\documentclass[fleqn,usenatbib]{mnras}

\usepackage{newtxtext,newtxmath}

\usepackage[T1]{fontenc}
\usepackage{ae,aecompl}

\usepackage{graphicx}	
\usepackage{amsmath}	
\usepackage{amssymb}	
\usepackage{bm}
\usepackage{adjustbox}

\usepackage{hyperref}

\def\apjl{ApjL}

\newcommand{\ex}{\hat{\bm{x}}}
\newcommand{\ey}{\hat{\bm{y}}}
\newcommand{\ez}{\hat{\bm{z}}}
\newcommand{\bh}{\bm{\hat{b}}}
\newcommand{\bb}{\bm{\hat{b}\hat{b}}}
\newcommand{\bbdu}{\bm{\hat{b}\hat{b}: \nabla u}}
\newcommand{\bbdU}{\bm{\hat{b}\hat{b} :\nabla U}}
\newcommand{\bbdUs}{\bm{\hat{b}\hat{b}: \nabla U_0}}
\newcommand{\re}{\mathrm{Re}}
\newcommand{\Rem}{\mathrm{Re_M}}
\newcommand{\Pm}{\mathrm{Pm}}
\newcommand{\Reb}{\mathrm{Re_B}}
\newcommand{\Ref}{\mathrm{Re_4}}
\newcommand{\Remf}{\mathrm{Re_{M,4}}}
\newcommand{\Tr}{\alpha_{\rm Re}}
\newcommand{\Tm}{\alpha_{\rm M}}
\newcommand{\Ta}{\alpha_{\rm A}}
\newcommand{\ratio}{\Ta / \Tm}
\newcommand{\dpBm} {\langle 4\pi \Delta p / B^2  \rangle}
\newcommand{\dpmBm}{4 \pi \langle \Delta p \rangle / \langle B^2  \rangle}
\newcommand{\Cal}{\int d^3 \bm{x} \ }
\newcommand{\nub}{\mu_{\rm B}}
\newcommand{\Dp}{\Delta p}
\newcommand{\dpm}{\langle \Dp\rangle}
\newcommand{\dpB}{4\pi \Dp/ B^2}
\newcommand{\prel}{3 \nub \bbdU 4\pi/ B^2}
\newcommand{\orbit}{t \Omega}
\newcommand{\Ha}{D_{\nub} / D}

\title[MRI turbulence in weakly collisional discs]{Shearing-box simulations of MRI-driven turbulence in weakly collisional accretion discs}

\author[Kempski, Quataert, Squire \& Kunz]{
Philipp Kempski,$^{1}$\thanks{E-mail: philipp.kempski@berkeley.edu}
Eliot Quataert,$^{1}$
Jonathan Squire,$^{2}$
Matthew W. Kunz$^{3,4}$
\\
$^{1}$Department of Astronomy and Theoretical Astrophysics Center, University of California, Berkeley, Berkeley, CA 94720, USA\\
$^{2}$Department of Physics, University of Otago, 730 Cumberland St, North Dunedin, Dunedin 9016, New Zealand\\
$^{3}$Department of Astrophysical Sciences, Princeton University, 4 Ivy Lane, Princeton, New Jersey 08544, USA \\
$^{4}$Princeton Plasma Physics Laboratory, PO Box 451, Princeton, New Jersey 08543, USA}
\date{Accepted XXX. Received YYY; in original form ZZZ}

\pubyear{2019}

\begin{document}
\label{firstpage}
\pagerange{\pageref{firstpage}--\pageref{lastpage}}
\maketitle

\begin{abstract}
We present a systematic shearing-box investigation of MRI-driven turbulence in a weakly collisional plasma by including the effects of an anisotropic pressure stress, i.e. anisotropic (Braginskii) viscosity. We constrain the pressure anisotropy ($\Dp$) to lie within the stability bounds that would be otherwise imposed by kinetic microinstabilities. We explore a broad region of parameter space by considering different Reynolds numbers and magnetic-field configurations, including net vertical flux, net toroidal-vertical flux and zero net flux. Remarkably, we find that the level of turbulence and  angular-momentum transport are not greatly affected by large anisotropic viscosities: the Maxwell and Reynolds stresses do not differ much from the MHD result. Angular-momentum transport in Braginskii MHD still depends strongly on \textit{isotropic} dissipation, e.g., the isotropic magnetic Prandtl number, even when the anisotropic viscosity is orders of magnitude larger than the isotropic diffusivities. Braginskii viscosity nevertheless changes the flow structure, rearranging the turbulence to largely counter the parallel rate of strain from the background shear. We also show that the volume-averaged pressure anisotropy and anisotropic viscous transport decrease with increasing isotropic Reynolds number ($\re$); e.g., in simulations with net vertical field, the ratio of anisotropic to Maxwell stress ($\ratio$) decreases from $\sim 0.5$ to $\sim 0.1$ as we move from $\re \sim 10^3$ to $\re \sim 10^4$, while $\dpBm \rightarrow 0$. Anisotropic transport may thus become negligible at high $\re$.  Anisotropic viscosity nevertheless becomes the dominant source of heating at large $\re$, accounting for $\gtrsim 50 \%$ of the plasma heating. We conclude by briefly discussing the implications of our results for RIAFs onto black holes.
\end{abstract}




\section{Introduction}
\label{sec:intro}
Magnetohydrodynamic (MHD) turbulence driven by the magnetorotational instability (MRI; \citealt{bh91}) is widely considered to be one of the key engines powering angular-momentum transport in accretion discs. As a result, the growth of the MRI and the subsequent MHD turbulence it produces have been studied extensively over the years (see e.g. \citealt{hgb95}; \citealt{bh98};  \citealt{hbs01}).

One uncertainty in the application of the MRI is that the MHD fluid approximation is not well justified in a number of astrophysical systems. This includes radiatively inefficient accretion flows (RIAFs), in which the Coulomb mean free path is larger than the typical system size (\citealt{mq97}). Departures from the ideal-MHD framework are therefore required and at first glance it may seem necessary to model the system as a fully collisionless plasma in six-dimensional phase space. The goal of this paper is to better understand the nonlinear evolution of the MRI under such conditions.

While it has recently become possible to run kinetic simulations of the MRI using particle-in-cell codes (\citealt{rqv15}; \citealt{h15}; \citealt{ksq16};  \citealt{iglfs18}), such simulations remain far too expensive (at least in three dimensions) to explore parameter space. However, a variety of recent theory (\citealt{sckrh08}; \citealt{kss14}; \citealt{sn15}; \citealt{rqv15}) suggests that ion-Larmor scale kinetic instabilities such as the mirror and firehose instabilities, which grow readily in low collisionality, weakly magnetized plasmas, act to increase the effective collision rate via wave-particle interactions. This result is of great utility, as it at least partially motivates modeling the system as a weakly collisional plasma. The advantage of this framework is that non-ideal effects are simply introduced as additional terms in the ideal fluid equations, which is much simpler than evolving a six-dimensional distribution function of the plasma particles. 

The linear growth of the MRI differs from its ideal-MHD counterpart in both a collisionless (\citealt{qdh02}) and a weakly collisional (\citealt{b04}) plasma. However, despite some work with simplified fluid models and kinetic simulations, we lack detailed understanding of how kinetic physics affects the saturated MRI turbulence. Some insight has been gained by the work of \cite{sqk17b}, who focused on the nonlinear growth phase of the MRI in high-$\beta$ low-collisionality plasmas. They argued that due to the onset of the aforementioned  microinstabilities, the nonlinear growth phase of the kinetic MRI (KMRI) always returns to MHD-like evolution. Similarly, the saturation into turbulence   appeared to be unaffected by non-ideal physics. These results provided insight into the physics behind earlier work on collisionless accretion discs by \cite{shqs06}, who found that the  properties of KMRI-induced turbulence were not too different from MHD. This resemblance has also been found in global general-relativistic simulations (\citealt{fcgq15}; \citealt{fcgqt17}), which employed an extended-MHD framework with anisotropic viscosity and conductivity.       

In this paper, we carry out a systematic shearing-box study of MRI-induced turbulence in a low-collisionality plasma with explicit resistivity and viscosity. To model non-ideal effects we use Braginskii's closure for magnetized, weakly collisional plasmas (\citealt{br65}), commonly referred to as ``Braginskii MHD''. As explained below, the anisotropic viscosity in Braginskii's closure is equivalent to including an anisotropic pressure stress in the MHD equations. The closure is the simplest, well motivated model to capture key aspects of kinetic physics on large scales. 

There are a number of questions that motivate such a parameter exploration. Perhaps most importantly, it will clarify the relevance of non-ideal physics for angular-momentum transport and plasma heating, including the additional contribution to the total stress tensor that comes directly from the pressure anisotropy. 

While the Maxwell and Reynolds stresses have been studied extensively in MHD, significantly less is known about the non-ideal, anisotropic viscous stress. Most simulations to date found that its contribution to angular-momentum transport is smaller than, but comparable to, the Maxwell stress. However, given that the pressure anisotropy is driven by gradients in the velocity field, we may expect isotropic dissipation to influence the anisotropic transport. It is therefore instructive to look at the relationship between anisotropic stress and the dimensionless isotropic Reynolds numbers.      

Exploring a range of isotropic viscosities and resistivities is vital for a second reason. One of the most striking results of previous work on MRI-generated turbulence is the  dependence on isotropic dissipation. \cite{ll07} and  \cite{fplh07} showed that the MRI saturation amplitude is very sensitive to the choice of  viscosity and resistivity, with a particularly strong dependence on their ratio, the magnetic Prandtl number.
It is plausible to speculate that an additional large anisotropic viscosity may alter the effective Prandtl number. This claim is further motivated by the kinetic MRI simulations of \cite{ksq16}, who showed that a high-$\beta$ collisionless plasma behaved in some ways like a high-magnetic-Prandtl-number fluid. It is therefore unclear to what extent we should expect to recover the usual Prandtl-number dependence in Braginskii MHD.

Another important factor to consider is how the very building blocks of MHD turbulence are modified in low-collisionality plasmas and whether this may change the large-scale turbulent state in low-collisionality accretion discs. \cite{sqs16} and \cite{skqs17a} showed that collisionless and weakly collisional plasmas cannot support linearly polarized shear-Alfv\'en waves above a critical amplitude due to a cancellation between the Lorentz force  and the anisotropic-pressure force. It is unclear if and how this might affect the large-scale turbulent properties in accretion discs.

The layout of this paper is as follows. In Section \ref{sec:method} we discuss the method and setup for our study of turbulence in Braginskii MHD. The main focus is on boxes threaded by a net vertical magnetic field, with the corresponding results described in Section \ref{sec:vertical}. We consider other initial field configurations in Section \ref{sec:otherfield}. Finally, Section \ref{sec:dis} summarizes our key results and discusses current limitations and future directions. 

\section{Method}
\label{sec:method}
\subsection{Equations} \label{sec:equations}
We use the  pseudo-spectral code SNOOPY (\citealt{ll07}) to evolve the incompressible MHD  equations with anisotropic pressure in a shearing box:
\begin{gather}
\bm{\nabla \cdot U}=0,  \label{eq:continuity} \\
\begin{aligned}
\frac{D\bm{U}}{Dt} = &-2\bm{\Omega \times U} + 2\Omega S x \ex  -\bm{\nabla} \left( p_\perp + \frac{B^2}{8\pi} \right)  \\ &
 + \bm{\nabla \cdot} \left[ \bb \left( \frac{B^2}{4\pi} + \Dp\right) \right]  +\nu \nabla ^2 \bm{U}, \\ \label{eq:momentum}
\end{aligned} \\
\frac{D\bm{B}}{Dt} =\bm{B \cdot} \bm{\nabla \bm{U}} +\eta \nabla ^2 \bm{B},   \label{eq:induction}
\end{gather}
where $\bh = \bm{B}/B$ is the unit vector along the magnetic field $\bm{B}$ and the density has been set to unity. Because our model is incompressible, we choose $p_\perp$ at each timestep so as to satisfy $\bm{\nabla \cdot U}=0$. We include an explicit isotropic viscosity $\nu$ and resistivity $\eta$. The velocity field $\bm{U}$ consists of a background shear $\bm{U_0}$ and perturbations $\bm{u}$: $\bm{U} = \bm{U_0} + \bm{u}$. We adopt an equilibrium Keplerian background profile $\bm{U_0} = -S x \ey$, with $S = \frac{3}{2} \Omega$, and we use the code to compute the evolution of $\bm{u}$ and $\bm{B}$. We use the $2/3$ de-aliasing rule to prevent spurious modes originating from the nonlinear terms in \eqref{eq:continuity}--\eqref{eq:induction}.

At any timestep, the pressure anisotropy $\Dp$ entering the momentum equation \eqref{eq:momentum} is calculated via:
\begin{equation} \label{eq:delp}
\begin{aligned}
\Dp= p_{\perp} - p_{\parallel} & =  3 \nub \bbdU, \\
\end{aligned}
\end{equation}
where $p_{\perp}$ and $p_{\parallel}$ are the thermal pressures in the directions perpendicular and parallel to the local magnetic field (\citealt{br65}). Equation \eqref{eq:delp} can be obtained from the kinetic evolution equations of the plasma (\citealt{cgl}; \citealt{k83}; \citealt{scrr10}) in the weakly collisional regime $\nu_c / |\bm{\nabla u}|  \gg 1$, where $\nu_c$ is the collision rate of the plasma. In equation \eqref{eq:delp} we also assume that the effect of heat fluxes on the pressure anisotropy can be neglected, an assumption that is formally valid when $ \nu_c / |\bm{\nabla u}|  \gg \beta^{1/2}$ (\citealt{mt71}), where $\beta=8 \pi p / B^2$ is the ratio of thermal to magnetic pressure (see \citealt{sqk17b} for more discussion of the different regimes). 

Upon substitution into \eqref{eq:momentum}, $\Dp$ can be shown to behave as a diffusion operator acting along $\bh$. Its role is therefore to damp the component of the velocity along the local magnetic field that has gradients along the local magnetic field. Due to its diffusive contribution, throughout this work we will refer to the coefficient $\nub$ as anisotropic (or Braginskii) viscosity.

\subsection{Modeling Kinetic Microinstabilities} \label{sec: subgrid}

The background shear and MRI-induced magnetic-field growth naturally lead to a  finite pressure anisotropy. However,  once $\Dp$ becomes comparable to the magnetic pressure, plasma instabilities are excited that are not fully captured by the set of equations \eqref{eq:continuity}--\eqref{eq:induction}. The two main microinstabilities that need to be accounted for are the mirror instability (\citealt{b66}; \citealt{h69}), excited if
\begin{equation} \label{eq:mirror}
\centering
\Dp\gtrsim \frac{B^2}{ 8 \pi}
\end{equation}
and the firehose instability (\citealt{r56}; \citealt{c58}; \citealt{p58}), which is excited when
\begin{equation} \label{eq:firehose}
\Dp\lesssim -\frac{B^2}{4 \pi}.
\end{equation}
Wave-particle interactions induced by these instabilities increase the effective collisionality of the system, which in turn acts to isotropize the pressure tensor. As a result, the mirror/firehose instability, excited by a growing/declining pressure anisotropy, acts to halt further growth/decline. Kinetic simulations have shown that these instabilities tend to pin the anisotropy near the instability thresholds (\citealt{ksq16}). 

We include this kinetic result by imposing hard-wall limits on $\Dp$. If $\Dp$ is driven outside of the stability limits given by $-B^2 /4\pi < \Dp< B^2 / 8\pi$, then it is  pinned to either $\Dp= -B^2/4\pi$  or $\Dp= B^2 /8\pi$, depending on which boundary is crossed. Otherwise, $\Dp$ is determined by equation $\eqref{eq:delp}$.

By using instantaneous bounds on $\Dp$, we essentially assume that the only effect of microinstabilities is to halt the growth of $\Dp$, with no direct change to other fluid quantities. For the firehose instability, this limiting behaviour arises due to particle scattering, while for the mirror instability, there is a long phase where small-scale magnetic fluctuations grow secularly in time. Although the limiter model can, in principle, capture the effect of either scattering or secular growth if we consider $\bm{u}$ and $\bm{B}$ in equations \eqref{eq:continuity}--\eqref{eq:induction} to be large-scale averages, there may be other poorly understood effects that are not captured. In addition, the assumption that $\Dp$ limiters act instantaneously is incorrect for motions with timescales approaching the ion gyro time. Because we find that the cascade continues almost unaffected to scales well below the Braginskii viscous scale, this could mean that the smallest-scale motions will be more strongly affected by $\Dp$ forces than we assume here. It is unclear if and how these additional effects could be incorporated in a fluid model, and a kinetic description is likely necessary to fully capture all the underlying physics (see \citealt{sckrh08}, \citealt{kss14}, \citealt{sqk17b} for more discussion).                   

We have also run two simulations without limiters at the firehose instability threshold. These no-firehose-limiter simulations are partially justified by the fact that the parallel firehose instability is already present in the Braginskii equations (by contrast, the mirror instability is not). However, the kinetic simulations in \cite{kss14} showed that the oblique firehose instability is probably more important for maintaining $\Dp$ near the instability threshold. For this reason, most of our simulations have both firehose and mirror limiters included.   

\subsection{Setup} \label{sec:setup}
Throughout this work we set $\Omega = 1$. Since we want to explore any pressure-anisotropy-induced differences, for every Braginskii MHD simulation we also carry out a corresponding MHD simulation in which $\Dp= 0$. Given the sensitivity of MRI turbulence to isotropic dissipation, we test a number isotropic viscosities $\nu$ and resistivities $\eta$. We define the associated dimensionless Reynolds number,
\begin{equation}
\re = \frac{S L_z^2}{\nu};
\end{equation}
the magnetic Reynolds number, 
\begin{equation}
\Rem = \frac{S L_z^2}{\eta};
\end{equation}
and their ratio, the magnetic Prandtl number,
\begin{equation}
\Pm = \frac{\nu}{\eta}.
\end{equation}
We also define the analogous Braginskii Reynolds number,
\begin{equation}
\Reb = \frac{S L_z^2}{\nub}.
\end{equation}

To quantify the turbulent angular-momentum transport, we define the dimensionless transport coefficient
\begin{equation}
\alpha = \Tr + \Tm  + \Ta ,
\end{equation}
where 
\begin{equation}
\Tr =  \langle v_x v_y \rangle \ / \ \big(S^2 L_z ^2\big),  
\end{equation}
\begin{equation}
\Tm = - \Big \langle \frac{B_x B_y}{4\pi} \Big \rangle \ / \  \big( S^2 L_z ^2 \big), 
\end{equation}
\begin{equation}
\Ta =  - \Big \langle \frac{\Dp}{B^2} B_x B_y \Big \rangle \ / \ \big( S^2 L_z^2 \big) 
\end{equation}
are the contributions from the volume-averaged ($\langle ... \rangle$) Reynolds stress, Maxwell stress and anisotropic viscous stress respectively, normalized by $S^2 L_z ^2$. This is the incompressible version of the compressible transport parameter, which is usually normalized using the initial pressure (see e.g. \citealt{hgb95}; \citealt{shqs06}).  While $\Tm$ and $\Tr$ are present both in ordinary MHD and in Braginskii MHD, $\Ta$ requires a pressure anisotropy and is therefore only nonzero in Braginskii MHD. 

To capture the most important parasitic modes that break up the MRI ``channel'' modes into turbulence, most of our simulations are in horizontally elongated boxes of size $L_x = 4$, $L_y = 4$ and $L_z = 1$ (\citealt{bmcrf08}; \citealt{pg09}; \citealt{ll10}). However, we did also explore other aspect ratios. We find that Braginskii MHD results are particularly sensitive to box size, with dramatically different results in horizontally narrow boxes with net vertical flux (see Appendix \ref{sec:box}). 

In order to satisfy the Courant condition at large $\nub$, we are limited to rather modest resolutions by current standards, despite sub-cycling over the $\Dp$ term in equation \eqref{eq:momentum} in simulations with large $\nub$ (where we used 5 or 8 as the maximum number of sub-cycles per main MHD timestep). Most of our full simulations in $4 \times 4 \times 1$ boxes have resolution $256 \times 128 \times 64$. To test very large $\nub$,  we also ran a number of lower resolution simulations. In each case, the initial magnetic field and background shear are perturbed with small-amplitude white noise. 

The isotropic viscosities and resistivities are chosen such that the dissipative scales are properly resolved. This places an upper bound on the $\re$ and $\Rem$ that we can explore in a full Braginskii simulation, given the attainable resolutions. To explore the relationship between anisotropic stress and the isotropic diffusivities, ideally we would like to cover a broad range of viscosities and resistivites, and explore the limit $\re \rightarrow \infty$, $\Rem \rightarrow \infty$ at fixed $\Pm$. However, the required resolutions are computationally unfeasible with the numerical methods that we use here.

To test larger Reynolds numbers in higher resolution, we perform a number of   ``Composite'' MHD--Braginskii MHD simulations. In these simulations, we first evolve the equations of MHD for a time $\orbit = 100$. The turbulent MHD flow fields are then restarted with anisotropic viscosity included and evolved further for several $\Omega^{-1}$. These composite simulations are motivated by our observation that anisotropic viscosity transforms MHD flow fields into Braginskii-like flow fields on timescales shorter than the orbital time.
We have tested that, by restarting MHD turbulence with anisotropic viscosity, we are able to recover the typical $\Dp$ and $\Ta$ of the corresponding full Braginskii simulation in a fraction of $\Omega^{-1}$. We show an example of this behavior in Appendix \ref{sec:testmixed}. 

This method offers insight into the statistics of Braginskii MHD turbulence at large Reynolds numbers, even when the Braginskii equations are evolved for a rather short amount of time. As a result, the composite simulations enable us to explore isotropic Reynolds numbers ($\re \sim 10^4$) and resolutions ($768 \times 384 \times 192$ and $384 \times 192 \times 96$) that are otherwise unattainable for a full Braginskii simulation with large $\nub$.  

We also perform a number of simulations in which the viscosity and resistivity are replaced by hyperdiffusion operators, $\nu_4 \nabla^4 \bm{ U}$ and $\eta_4 \nabla^4 \bm{B}$. We define the associated dimensionless quantities $\Ref = S L_z^4/\nu_4$ and $\Remf = S L_z^4/\eta_4$. Using hyperdiffusion serves as an alternative method to probe larger effective Reynolds numbers, by potentially increasing the size of the inertial range without increasing the resolution. The $k^4$ dependence of hyperdiffusion allows us to dissipate energy above the grid scale even for very small $\nu_4$, without constraining the turbulence at intermediate wavenumbers, thus mimicking turbulence at somewhat larger $\re$ than would be possible with standard diffusion operators (note, however, that the $\Ref$ defined above is \textit{not} the effective Reynolds number of our hyperdiffusion simulations).

While our main focus is on simulation domains with a net vertical field (Section \ref{sec:vertical}), we also discuss other initial magnetic-field configurations: net vertical and toroidal field (\ref{sec:toroid}), as well as zero net flux (\ref{sec:zero}).

\section{Net Vertical Field }
\label{sec:vertical}
The main part of our study concerns boxes initially threaded by a purely vertical magnetic field, 
\begin{equation}
\langle \bm{B} \rangle = B_0 \ez.
\end{equation}
We choose $B_0 = \sqrt{8\pi/1348} \ \Omega L_z$, so that the fastest-growing MRI mode in MHD has wavelength $\lambda_{\rm MRI}= 0.25 L_z$. 

Table \ref{tab:vertical} gives a summary of our different choices of $\re$, $\Rem$ and $\Reb$. The main result of this section concerns anisotropic stress and its dependence on the values of $\re$ and $\Rem$. However, we postpone our discussion of this result until section \ref{sec:alphaa} and first look at the overall Braginskii MHD evolution, and its similarities and differences relative to MHD. 

\begin{table*}
	\centering
\caption{Summary of simulations with net vertical field. \textbf{Full simulations:} each simulation set at a fixed $\re$, $\Rem$ consists of an MHD simulation (top) and Braginskii MHD simulation(s) (bottom). The transport coefficients $\alpha$ and the mean pressure anisotropies were averaged over $\orbit = 100 - 200$. ``Braginskii$^{***}$" indicates a Braginskii MHD simulation without firehose limiter included. \textbf{Composite MHD--Braginskii MHD simulations:} MHD fields are restarted at $\orbit = 100$ using Braginskii MHD. For simulations at resolution $384 \times 192 \times 96$,  averages were taken over $\orbit = 101-110$. For the simulations at higher resolution, averages are over $\orbit = 101-102$. ``Composite$^{***}$" indicates a simulation where MHD flow fields were restarted in Braginskii MHD without firehose limiter. \label{tab:vertical}  }
  \begin{adjustbox}{width=\textwidth}

\begin{tabular}{cccccccccccccc}\hline\hline
Sim. Type & Resolution & $\re$ & $\Rem$ & $\Pm$ &  $\Reb$ & $\Tr$ & $\Tm$ & $\Ta$ & $\alpha$ & $\dpmBm$ &$\dpBm$ & $\ratio$  \\
\hline\hline
\\
Full MHD & $(256,128,64)$ & $6000$ & 3000 & 0.5& -- & 0.0065   & 0.031  &  -- &  0.037 & --  & -- &--  \\
Full Braginskii & $(256,128,64)$ & $6000$ & 3000 &  0.5 & 0.75 & 0.0084   & 0.031  &  0.0086 & 0.048& 0.21  & 0.15 & 0.28   \\
\\
Full MHD & $(256,128,64)$ & $750$ & 750 & 1 & -- &  0.0051  & 0.020   &  -- &  0.025 & -- & -- & --   \\
Full Braginskii & $(256,128,64)$ & $750$ & 750 & 1 & 0.75 &  0.0065  & 0.019    & 0.0090   & 0.035 & 0.41 &  0.30 &0.47\\
\\
Full MHD & $(256,128,64)$ & $4500$ & 4500 & 1 & -- & 0.0062   & 0.036  &  -- &  0.043 & -- &  -- &-- \\
Full Braginskii & $(256,128,64)$ & $4500$ & 4500 & 1 & 0.75 & 0.0097    & 0.044   &  0.011 & 0.065 & 0.19  & 0.12 &0.25 \\
\\
Full MHD & $(256,128,64)$ & $1500$ & 3000 & 2 & -- & 0.0078   &0.046   &  -- & 0.054 & --  & -- & --   \\
Full Braginskii & $(256,128,64)$ & $1500$ & 3000 & 2 & 0.75 & 0.012   & 0.049   & 0.016  &    0.077 & 0.28  & 0.19 & 0.35\\
Full Braginskii$^{***}$ & $(256,128,64)$ & $1500$ & 3000 & 2 & 0.75 & 0.011   & 0.045   & 0.016  &    0.072 & -0.021  & -22.3 & 0.36\\
\\
Full MHD & $(256,128,64)$ & $750$ & 6000 &8 & -- &  0.015   &  0.10  &  -- & 0.12 & -- & -- & --   \\
Full Braginskii & $(256,128,64)$ & $750$ & 6000 & 8 & 0.75 & 0.022  &  0.11  & 0.031  &  0.17 & 0.21  & 0.13 & 0.27\\
\hline
\\

Full MHD & $(192,96,48)$ & $1500$ & 750 & 0.5 & -- &0.0047  & 0.015 & -- & 0.019 & -- & -- & --   \\
Full Braginskii & $(192,96,48)$ & $1500$ & 750 & 0.5 & 0.3 & 0.0049 &0.014  &0.0062   & 0.025 & 0.39 & 0.29  &0.46  \\
\\
Full MHD & $(192,96,48)$ & $1500$ & 3000 & 2& -- & 0.0082  & 0.049  &  -- &0.057 & --  & -- & --   \\
Full Braginskii & $(192,96, 48)$ & $1500$ & 3000 & 2 & 300 & 0.0084 & 0.045    & 0.0006   &  0.054 & 0.012 & 0.032 & 0.013    \\
Full Braginskii & $(192,96, 48)$ & $1500$ & 3000 & 2 & 75 & 0.0090 & 0.050    & 0.0020  &  0.061 & 0.037 & 0.069 & 0.044    \\
Full Braginskii & $(192,96, 48)$ & $1500$ & 3000 & 2 & 15 & 0.0093 & 0.047    & 0.0065   &  0.063 & 0.12 & 0.14 & 0.15    \\
Full Braginskii & $(192,96, 48)$ & $1500$ & 3000 & 2 & 3 & 0.011 & 0.047    & 0.013   &  0.071 & 0.23 & 0.18 & 0.29    \\
Full Braginskii & $(192,96, 48)$ & $1500$ & 3000 & 2 & 0.3 & 0.0097 & 0.040    & 0.013   &  0.063 & 0.26 & 0.17 & 0.33    \\
\hline
\\
Full MHD & $(128,64,32)$ & $1500$ & 3000 & 2& -- & 0.0069  & 0.036  &  -- &0.043  & --  & -- & --   \\
Full Braginskii & $(128,64,32)$ & $1500$ & 3000 & 2& 0.075 & 0.011   &0.047   & 0.013  &0.072 & 0.21 & 0.13  &0.29   \\
\\
\hline
\hline
\\
Composite & $(384,192,96)$ & $750$ & 750 & 1 & 0.75 & 0.0043 & 0.011 & 0.0055 & 0.021 & 0.42 & 0.31 & 0.49   \\ \\
Composite &$(384, 192, 96)$ & $3000$ & 3000 & 1 & 0.75 & 0.0071  & 0.026 & 0.0090 & 0.042 & 0.28  & 0.19 & 0.35  \\ \\
Composite &$(384,192,96)$ & 10500 & 10500 & 1 & 0.75 & 0.011 & 0.057 & 0.011 & 0.078 & 0.12  &0.060 &  0.19   \\ 
Composite$^{***}$ &$(384,192,96)$ & 10500 & 10500 & 1 & 0.75 & 0.013 & 0.073 & 0.0079 & 0.094 & -0.28  & -17.7 &  0.11   \\
Composite &$(576,288,144)$ & 10500 & 10500 & 1 & 0.75 & 0.014 & 0.068 & 0.013  & 0.095  & 0.13  &0.070 &  0.19  \\ \\
Composite &$(768, 384, 192)$ & 21000 & 21000 & 1 & 0.75 & 0.010 & 0.081 & 0.0088 &  0.10 & 0.076  &0.026 & 0.11 \\ \\
Composite &$(768,384,192)$ & 10500 & 42000 & 4 & 0.75 & 0.020 & 0.12 & 0.017 & 0.15  & 0.076 & 0.016 & 0.14   \\
\\
\hline
\hline
\end{tabular}

\end{adjustbox}

\end{table*}

\subsection{Comparison to MHD}

\begin{figure}
\centering
 \includegraphics[scale = 0.58]{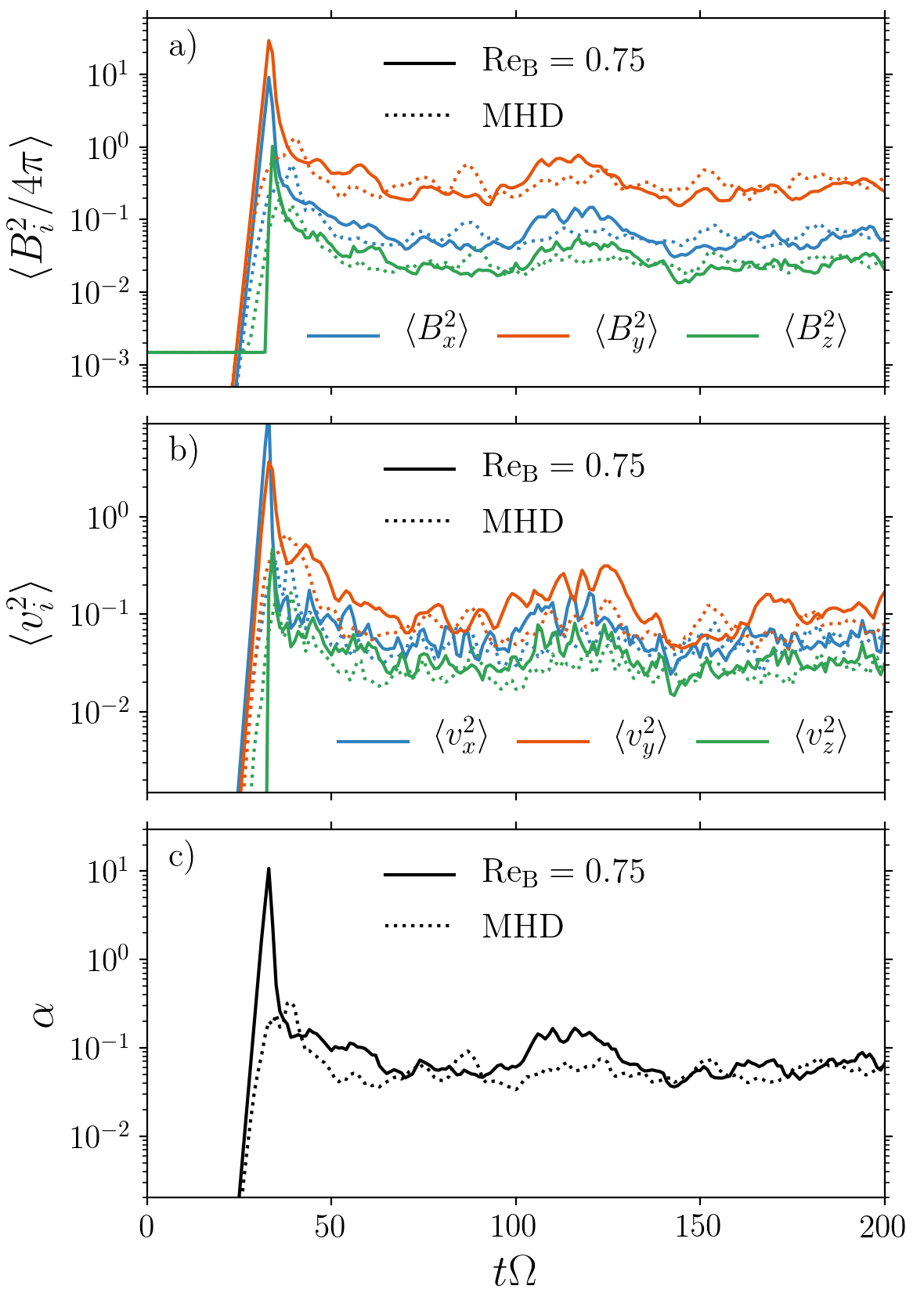}
   \caption{Evolution of energy densities and the transport coefficient $\alpha$ in simulations with $\re = 1500$ and $\Pm=2$ with net vertical flux. In the saturated phase the Braginskii model with $\Reb=0.75$ (solid lines) closely matches MHD evolution (dotted lines).  \label{fig:Re1000Pm2_evol} }
\end{figure}


In each of our simulations the qualitative evolution follows the same pattern. The initial small-amplitude perturbations are amplified by the MRI. The amplification continues until the MRI modes reach large amplitudes and become unstable to parasitic instabilities  (\citealt{gx94}). These are secondary instabilities that grow on the large field and flow gradients in the MRI mode, causing it to break up into turbulence.  \cite{sqk17b} argued that the dominant parasitic modes are not too different in Braginskii MHD. This is broadly consistent with our numerical results summarized below, although we do often see somewhat larger $\alpha$ in the Braginskii case during a short initial transient phase. In addition, we find a very significant box-size dependence in Braginskii MHD (see Appendix \ref{sec:box}).

Figure \ref{fig:Re1000Pm2_evol} shows the evolution of the simulation with $\re = 1500$ and $\Pm=2$. The dotted lines track the MHD evolution, while the solid lines correspond to Braginskii MHD with $\Reb = 0.75$. The two models show similar behavior and the resultant turbulent  energy densities and angular-momentum transport are close to identical. The main qualitative differences  appear at early times, during the MRI growth phase. Braginskii viscosity delays the growth along the initial field direction ($\langle B_z^2 \rangle$ and $\langle v_z^2 \rangle$) through stronger damping of the initial white-noise perturbations. In addition, in the Braginskii MHD simulation the MRI is able to grow to larger amplitudes, before it eventually saturates.

For all of the cases we simulated, the transport is not greatly affected by the anisotropic viscosity. This includes our simulations with both limiters included, as well as our full simulation without a firehose limiter. The main difference is the presence of the anisotropic viscous stress. The Maxwell and Reynolds stresses are very similar in both MHD and Braginskii MHD. 

\begin{figure} 
 \centering
 \includegraphics[scale = 0.5]{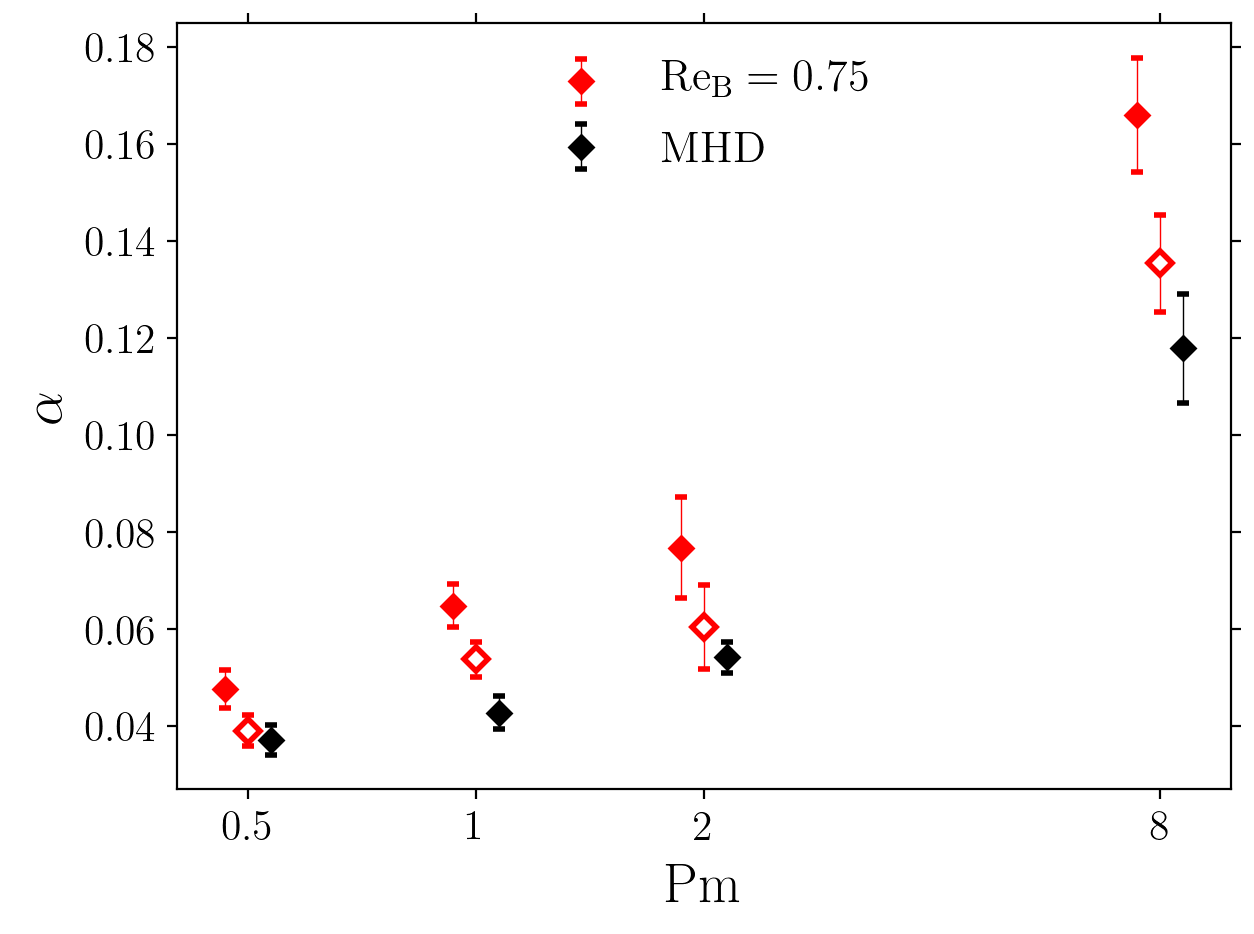}
   \caption{Temporal average of $\alpha$ for different $\Pm$ in MHD (black) and Braginskii MHD with $\Reb=0.75$ (red).  The filled diamonds represent the total transport coefficient $\alpha$. The empty red diamonds count the contribution from $\Tr+\Tm$ in the Braginskii simulation. The error bars are the standard deviations of the time-averaged $\alpha$. The Maxwell and Reynolds stresses in Braginskii MHD show the usual MHD-like Prandtl-number dependence; the difference between MHD and Braginskii MHD is caused primarily by anisotropic stress. The $\Pm=0.5$ and $\Pm=2$ simulations have $\Rem = 3000$, the $\Pm=1$ simulation has $\Rem = 4500$ and $\Pm=8$ has $\Rem=6000$. The points at a fixed $\Pm$ have been slightly displaced for visualization purposes. \label{fig:alphaPm} }
\end{figure}

The remarkable similarities between the MHD and weakly collisional solutions draw us to an interesting conclusion: even  with large anisotropic viscosity, the turbulent amplitudes are still set primarily  by the isotropic Reynolds and magnetic Reynolds numbers. This is illustrated in Figure ~\ref{fig:alphaPm}, where we show the time-averaged transport coefficients of the MHD (black) and Braginskii MHD (red) runs at different $\Pm$. For the Braginskii runs we also plot the transport due to just the Reynolds and Maxwell stresses as empty red diamonds. The error bars shown in Figure \ref{fig:alphaPm} are estimates for the standard deviations of the average transport coefficients, which use a binning time of $t = 10 \Omega^{-1}$ (due to the short Braginskii timesteps, our simulations were not evolved long enough for an error analysis similar to \citealt{ll10}). Nevertheless, they illustrate that the $\Tr + \Tm$ of the Braginskii calculation agrees well with MHD. In Braginskii MHD we recover the usual Prandtl-number dependence found in MHD, thus showing that  turbulence is still  strongly influenced by the isotropic diffusivities, even when $\Reb \ll \re$. 

The result that angular-momentum transport is not significantly affected by large anisotropic viscosities is partly due to the anisotropic-pressure limiters. These limit $\Dp$ to being comparable to the local field strength, which implies that the anisotropic stress cannot become significantly larger than the Maxwell stress. However, we see here that the box-averaged anisotropic stress can be substantially smaller than the Maxwell stress (e.g., $\Ta = 0.25 \Tm$ in the full Braginskii run with $\re=4500$ and $\Pm=1$; we also find that $\Ta \ll \Tm$ in high-$\re$ composite simulations, as discussed in Section \ref{sec:alphaa}), which is less obvious, and surprising in light of previous results (e.g., \citealt{shqs06} and \citealt{ksq16}, where $\Ta$ is comparable to $\Tm$). Perhaps even more surprising is that we find little dependence of the angular-momentum transport on anisotropic viscosity even though the effective ($\Dp$-limited) Braginskii viscosities considered here are much larger than the isotropic diffusivities. This is in contrast to the strong dependence of angular-momentum transport in a shearing box on isotropic diffusivities (\citealt{fplh07}; \citealt{ll07}). In addition, we show in Section \ref{sec:alphaa} that angular-momentum transport is similar to MHD despite the fact that the flow structure in Braginskii MHD is quite different from MHD (e.g., Figure \ref{fig:bbdu_transition}).

\subsection{Anisotropic Transport in Braginskii MHD}

\label{sec:alphaa}

\begin{figure*}
\centering
 \includegraphics[scale = 0.47]{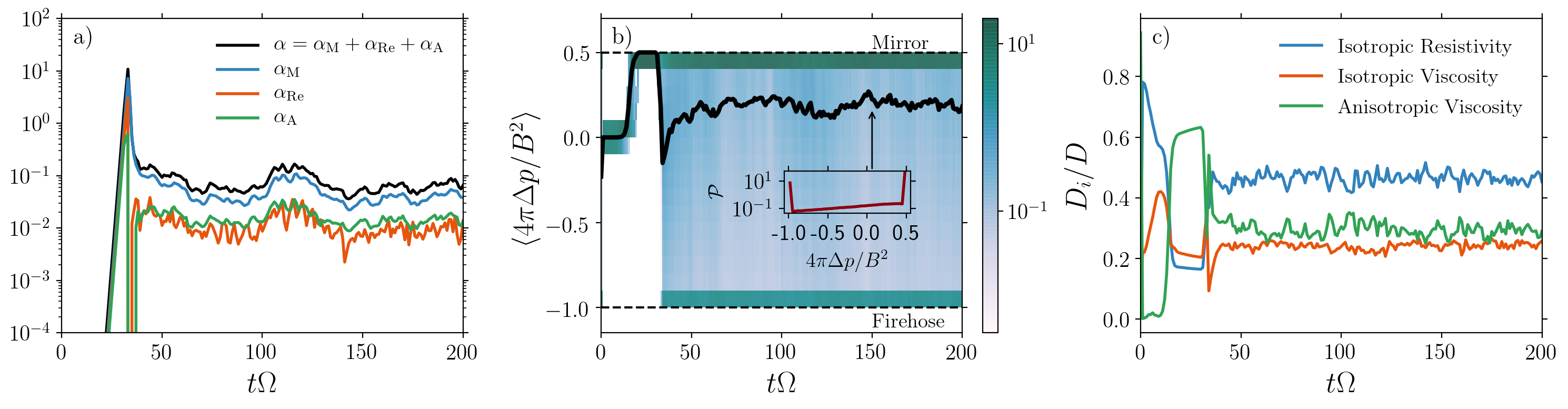}
   \caption{Braginskii MHD evolution for $\re=1500$, $\Pm=2$ and $\Reb=0.75$. \textbf{Panel a):} Evolution of the Maxwell (blue), Reynolds (orange), anisotropic (green) and total (black) transport coefficients in Braginskii MHD. \textbf{Panel b):} Evolution of the volume-averaged pressure anisotropy divided by (twice) the magnetic energy (black solid line). The background shading shows the underlying distribution of $\dpB$ over time, showing that the majority of cells lie on the mirror boundary (the width of the region near the mirror and firehose boundaries is exaggerated for visualization purposes). The inset shows this distribution across the simulation domain at the time $\orbit = 150$. \textbf{Panel c):} heating fractions of isotropic and anisotropic diffusivities over time. The heating is dominated by resistive heating (eq. \ref{eq:res_diss}), followed by anisotropic (eq. \ref{eq:anis_diss}) and isotropic (eq. \ref{eq:visc_diss}) viscous heating. See Figures \ref{fig:alpha_anis_vs_Re} -- \ref{fig:heating_vs_Re} for how these results depend on $\re$. \label{fig:alpha_breakdown}}
\end{figure*}

In this section we explore aspects of angular-momentum transport that are specific to Braginskii MHD. Having demonstrated that the Maxwell and Reynolds stresses tend to track their values in the complementary MHD simulations, our main focus is on the evolution of the anisotropic stress and pressure anisotropy.

In Figure \ref{fig:alpha_breakdown}a we show the evolution of the Maxwell, Reynolds and anisotropic viscous stresses for the simulation with $\re=1500$, $\Pm=2$ and $\Reb=0.75$. The overall transport is dominated by the Maxwell stress, followed by the anisotropic and Reynolds stresses. This is similar to the results of the kinetic MRI simulations in \cite{ksq16} and the global extended-MHD simulation in \cite{fcgqt17}. In all of our simulations we find that the Maxwell stress dominates. 
    
Figure \ref{fig:alpha_breakdown}b shows the evolution of the box-averaged pressure anisotropy. In the initial growth phase of the MRI, $\Dp$ grows steadily until all cells are pinned at the mirror boundary. In the turbulent phase it then shows small oscillations around a roughly constant value. The background coloring shows the underlying $\dpB$ distribution of cells over time. The inset shows this distribution across the simulation domain at a selected time ($\orbit = 150$). Most cells are pinned at the microinstability limits, the majority being on the mirror side, giving an overall positive pressure anisotropy.  

Figure \ref{fig:alpha_breakdown}c shows the evolution of heating due to isotropic and anisotropic diffusion, normalized by the total dissipation,
\begin{equation} \label{eq:res_tot}
    D = D_{\eta} +   D_{\nu} + D_{\nub},
\end{equation}
where $D_{\eta}$ is the resistive heating,
\begin{equation} \label{eq:res_diss}
    D_{\eta} = - \frac{\eta}{4\pi} \Cal \bm{B \cdot}  \nabla^2 \bm{B} ,
\end{equation}
$D_{\nu}$ is the isotropic viscous heating,
\begin{equation} \label{eq:visc_diss}
 D_{\nu} = - \nu \Cal \bm{u \cdot} \nabla^2 \bm{u },  
\end{equation}
and $D_{\nub}$ is the anisotropic viscous heating,
\begin{equation} \label{eq:anis_diss}
    D_{\nub} = \Cal \Dp \ \bm{\bbdU} .
\end{equation}
The pressure anisotropy $\Dp$ in equation \eqref{eq:anis_diss} is computed with mirror and firehose limiters included.
In this simulation with $\re=1500$, $\Pm=2$ and $\Reb=0.75$, resistive heating dominates, followed by anisotropic viscous heating and isotropic viscous heating.

\subsubsection{Dependence on $\Reb$} \label{sssec:vs_reb}

\begin{figure}
\centering
 \includegraphics[scale = 0.63]{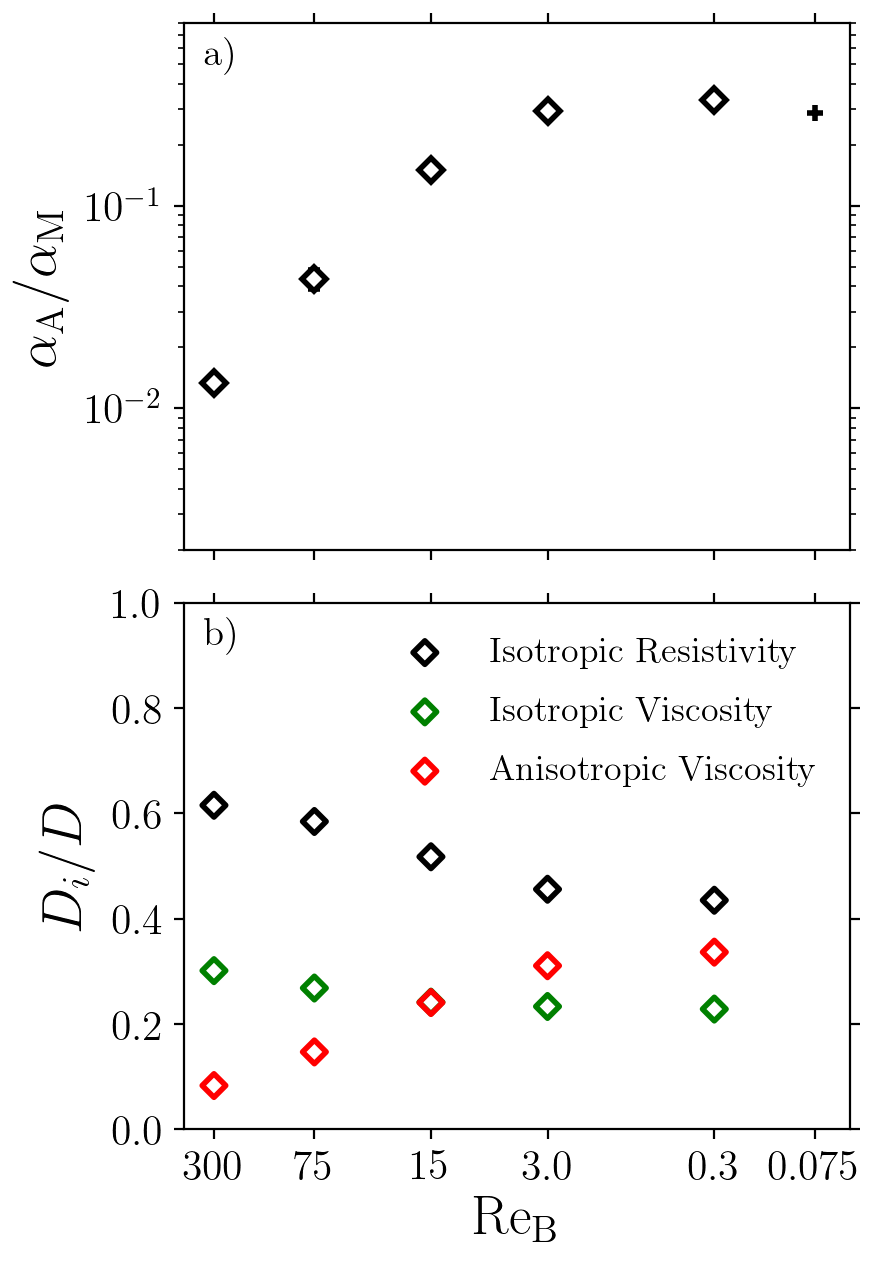}
   \caption{Dependence on Braginskii Reynolds number $\Reb$ in simulations with $\re=1500$, $\Pm=2$ at resolution $192 \times 96 \times 48$. \textbf{Panel a):} ratio of anisotropic to Maxwell stress, $\ratio$, as a function of Braginskii Reynolds number. The black ``+'' is our largest-$\nub$ test, performed at lower resolution ($128 \times 64 \times 32$). Error bars are plotted, but not visible, as they are smaller than the marker size. \textbf{Panel b):} temporal averages of heating fractions of isotropic and anisotropic diffusivities, as a function of $\Reb$. Anisotropic viscous heating increases with decreasing $\Reb$, until it becomes approximately constant at large anisotropic viscosities. The simulation with $\Reb=0.075$ at lower resolution is not shown, as it is uncertain whether the low resolution permits an accurate calculation of isotropic dissipation. \label{fig:alpha_nuB} }
\end{figure}

\begin{figure*}
\centering
 \includegraphics[scale = 0.65]{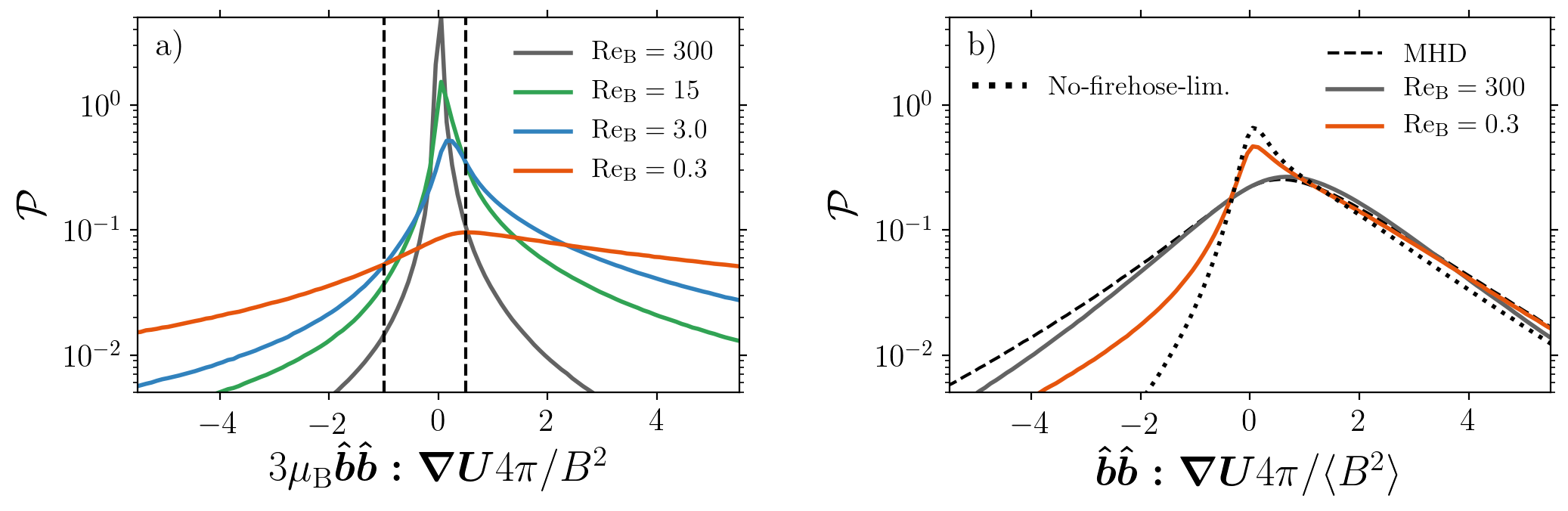}
   \caption{Distributions for the simulations summarized in Figure \ref{fig:alpha_nuB}. \textbf{Panel a):} The pre-limiter distributions of $ \dpB$ for different $\Reb$. The dashed vertical lines denote the firehose (left) and mirror (right) limits. $\ratio$ reaches a plateau when the distribution of $\prel$ (before limiters are applied) becomes wide compared to the hard-wall limits. \textbf{Panel b):} At small $\Reb$ there is a narrower distribution of $\bbdU 4\pi / \langle B^2 \rangle$. The dotted line is our no-firehose-limiter simulation with $\re=1500$, $\Pm=2$ and $\Reb=0.75$, in which large negative $\bbdU$ is even more suppressed (the distribution for $\Reb=0.75$ with firehose limiter included is very similar to the $\Reb=0.3$ distribution). These results are a consequence of anisotropic viscosity causing the turbulence to resist field-line stretching, i.e. $\bbdU$ is minimized.   \label{fig:hist_vs_nuB} }
\end{figure*}

One might expect that anisotropic viscous transport and anisotropic viscous heating will depend primarily on our choice of anisotropic viscosity $\nub$. We show the dependence on Braginskii Reynolds number for simulations with $\re=1500$ and $\Pm=2$ in Figure \ref{fig:alpha_nuB}. Figure \ref{fig:alpha_nuB}a shows that $\ratio$ increases with increasing anisotropic viscosity at large $\Reb$, reaching a plateau at small $\Reb$. Figure \ref{fig:alpha_nuB}b shows the dependence of the time-averaged heating fractions on $\Reb$. $\Ha$ has a dependence similar to $\ratio$, increasing with increasing $\nub$ at large $\Reb$ and approaching an approximately constant value at small $\Reb$. This is qualitatively similar to the results found by \cite{ssqk18} for strong Alfv\'enic turbulence. Note, however, that the final values of $\ratio$ and $\Ha$, reached at $\Reb \lesssim 1$, depend on $\re$ and $\Rem$ (see Figures \ref{fig:alpha_anis_vs_Re} and \ref{fig:heating_vs_Re}).

The plateaus in Figure \ref{fig:alpha_nuB} at small $\Reb$ (high $\nub$) are related to the fact that, due to the presence of limiters, the effective $\nub$ saturates. Anisotropic transport and heating are most sensitive to the choice of anisotropic viscosity at small $\nub$, when most fluid cells have $\Dp$ within the microinstability limits (eq. \ref{eq:mirror} \& \ref{eq:firehose}). But once $\nub$ is sufficiently large such that most cells lie outside of the limiter region, $\ratio$ and $\Ha$ reach a plateau. We illustrate this in Figure \ref{fig:hist_vs_nuB}a, where we show the distributions of $\prel$ for the different choices of $\Reb$. This is the pre-limiter $\Dp$ distribution divided by (twice) the magnetic energy, before any hard-wall limits are applied to $\Dp$ (the distribution with limiters is shown in the inset of Figure \ref{fig:alpha_breakdown}b). The anisotropic stress and anisotropic heating fraction reach an almost constant value once the pressure anisotropy distribution lies mostly outside of the dashed vertical lines denoting the mirror and firehose limits. We find that typically $\Reb \lesssim 3$ is enough to approach the asymptotic value. This can be explained as follows:  the presence of limiters causes the effective $\nub$ to saturate when $3 \nub \bbdUs = -3 \nub b_x b_y S \sim B^2 / 4\pi$, or equivalently when $\Reb \sim 4 \pi S^2 L_z^2 / B^2$. This gives $\Reb \sim$ a few for typical turbulent energy densities, which explains the plateaus in Figure \ref{fig:alpha_nuB}.

Figure \ref{fig:hist_vs_nuB}b shows that while the $\prel$ distribution becomes broader, the distribution of $\bbdU 4 \pi/ \langle B^2 \rangle$ becomes narrower with increasing $\nub$. In addition, large negative $\bbdU$ is more suppressed in simulations without firehose limiter. This can be explained by the work of \cite{ssqk18}, who demonstrated that anisotropic viscosity acts to minimize field-line stretching $\bbdU$ to resist changes in magnetic-field strength and make the flow ``magneto-immutable''.

\subsubsection{Dependence on $\re$ and $\Rem$: composite simulations} \label{sssec:vs_re}
Is isotropic dissipation important for the level of anisotropic transport? It is well known that the Maxwell and Reynolds stresses show a strong dependence on isotropic viscosity and resistivity. Given the interdependence of the pressure anisotropy and velocity-field gradients, it is plausible that $\Ta$ will also be sensitive to the choice of isotropic Reynolds numbers.  Our fully evolved Braginskii simulations (Table \ref{tab:vertical}) tentatively suggest that the relative importance of $\Ta$ is greater at low $\re$ and $\Rem$. This is best seen from the two simulations with $\Pm = 1$: $\ratio$ decreases from 0.47 for $\re = 750$ to 0.25 for $\re = 4500$, a change significantly larger than the characteristic temporal fluctuations. To explore this dependence in detail and cover a broad range of viscosities and resistivities, we first make use of our composite MHD--Braginskii MHD simulations (see Section \ref{sec:setup}) and then simulations with hyperdiffusion.

\begin{figure*}
\centering
 \includegraphics[scale = 0.47]{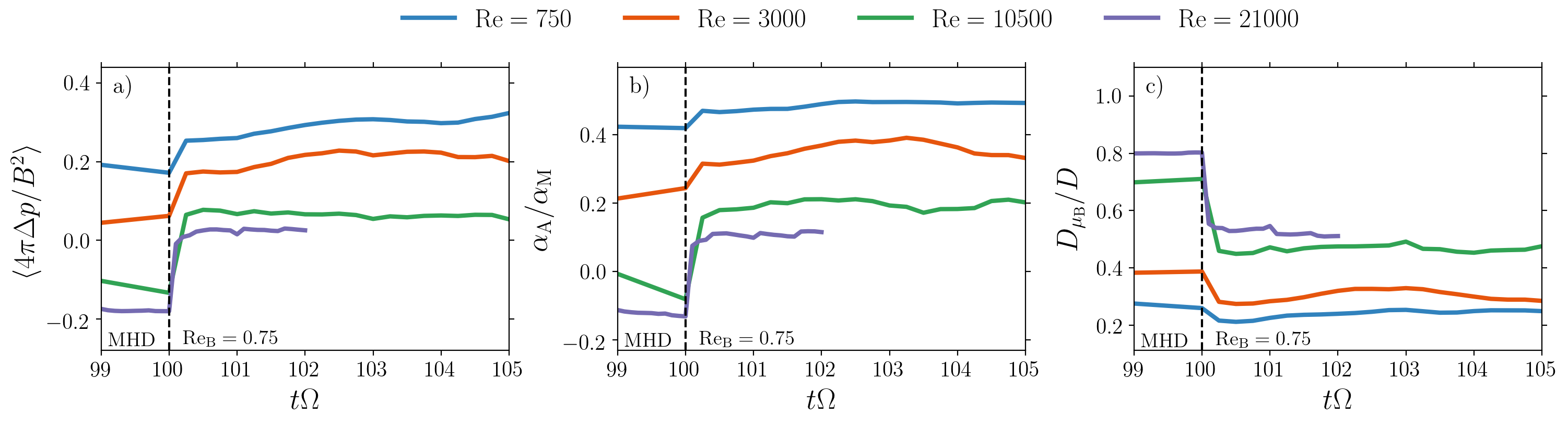}
   \caption{Evolution of a) anisotropic pressure, b) the ratio of anisotropic to Maxwell stress and c) the anisotropic viscous heating fraction in the composite MHD--Braginskii MHD simulations. The MHD turbulent flow field is restarted using Braginskii MHD with $\Reb = 0.75$ at $\orbit = 100$ (prior to this time $\Dp$, $\Ta$ and $\Ha$ are calculated using the MHD flow field with $\Reb = 0.75$ even though $\nub$ is not dynamically present in the MHD equations). Each color represents a different choice of isotropic Reynolds number with fixed $\Pm=1$. Anisotropic pressure and anisotropic transport decrease with increasing $\re$. Meanwhile, anisotropic viscous heating increases with increasing Reynolds numbers, accounting for $\gtrsim 50 \%$ of the total heating at large $\re$.    \label{fig:transition} }
\end{figure*}

\begin{figure*} 
\centering
 \includegraphics[scale = 0.6]{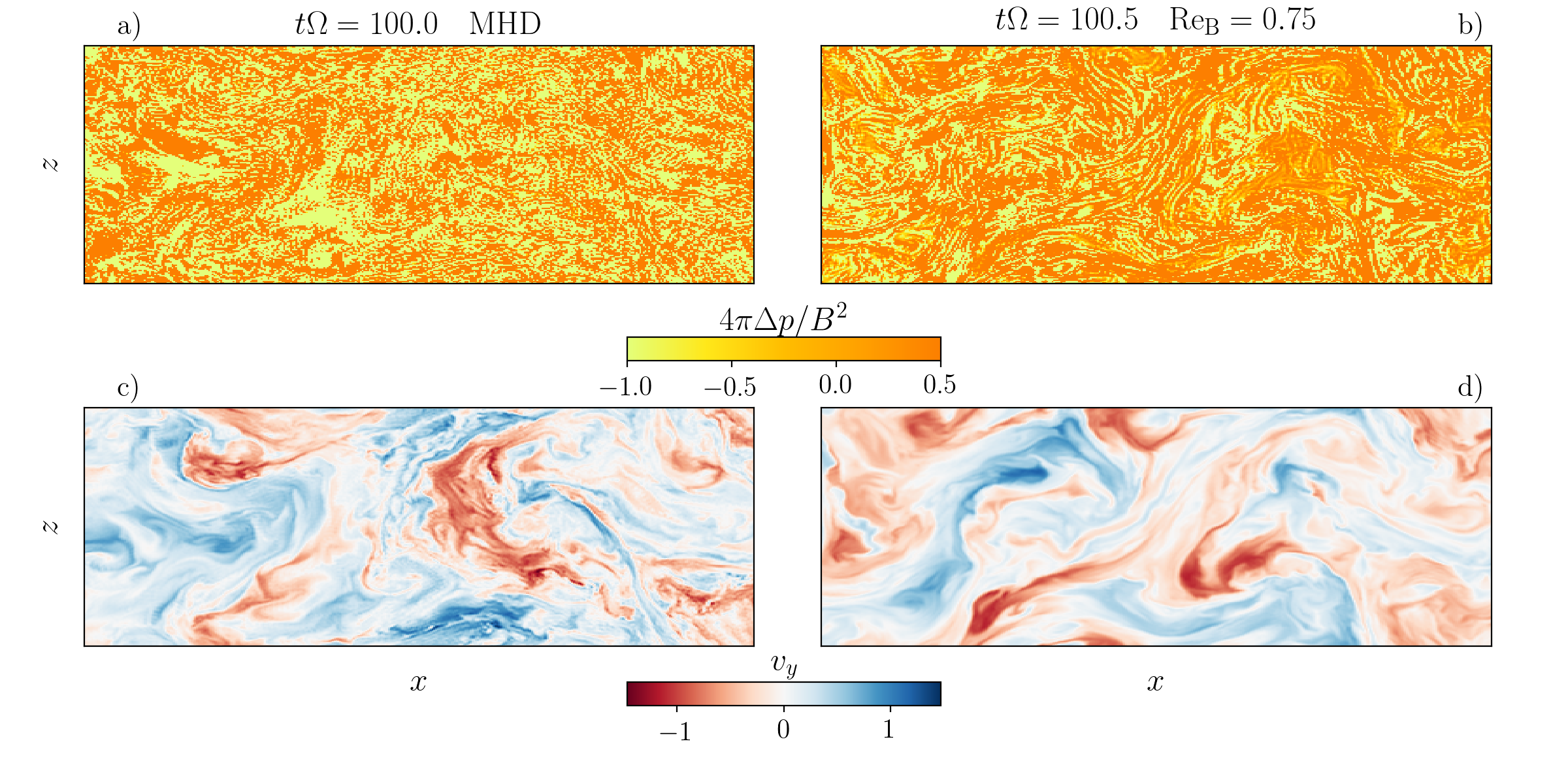}
   \caption{Snapshots from the composite simulation with $\re=21000$ and $\Pm=1$. \textbf{Top:} snapshots of $\dpBm$, before (left) and after (right) the MHD flow field is restarted using Braginskii MHD with $\Reb=0.75$ (the MHD $\Dp$ is calculated using the MHD flow field with $\Reb = 0.75$, even though $\nub$ is not present in the MHD equations). Most cells are pinned at the hard-wall limits, with numerous mirror-firehose neighboring cells in the MHD snapshot. These small-scale variations are smoothed by anisotropic pressure in the right panel. \textbf{Bottom:} smoothing of MHD small-scale velocity variations (left) after Braginskii viscosity is introduced (right). \label{fig:snap} }
\end{figure*}

\begin{figure} 
 \centering
 \includegraphics[scale = 0.6]{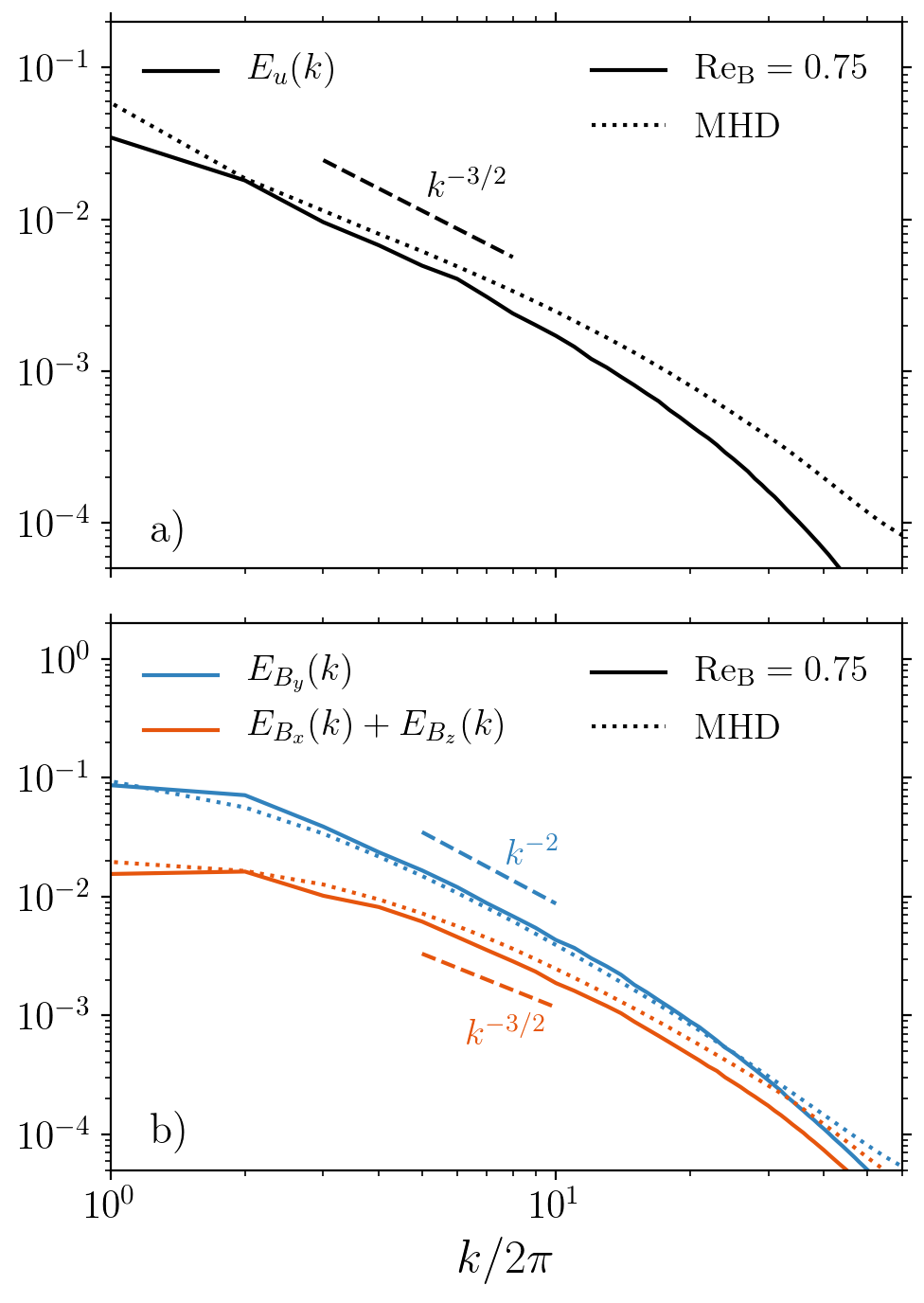}
   \caption{ Energy spectra for the composite simulation with $\re=21000$ and $\Pm=1$. The MHD part of the run ($t\Omega < 100$) is shown using dotted lines, while the Braginskii MHD part ($t\Omega>101$) with $\Reb = 0.75$ is shown as solid lines. \textbf{Panel a):} Braginskii MHD has a velocity-field spectral slope close to $-3/2$, which is slightly steeper than the corresponding MHD slope. There is extra damping of high-$k$ velocity fluctuations in the Braginskii case, due to the diffusive nature of the pressure anisotropy. \textbf{Panel b):}  Magnetic-energy spectra are not strongly affected by the presence of anisotropic viscosity and look very similar in MHD and Braginskii MHD.   \label{fig:spectra} }
\end{figure}

\begin{figure} 
\centering
 \includegraphics[scale=0.55]{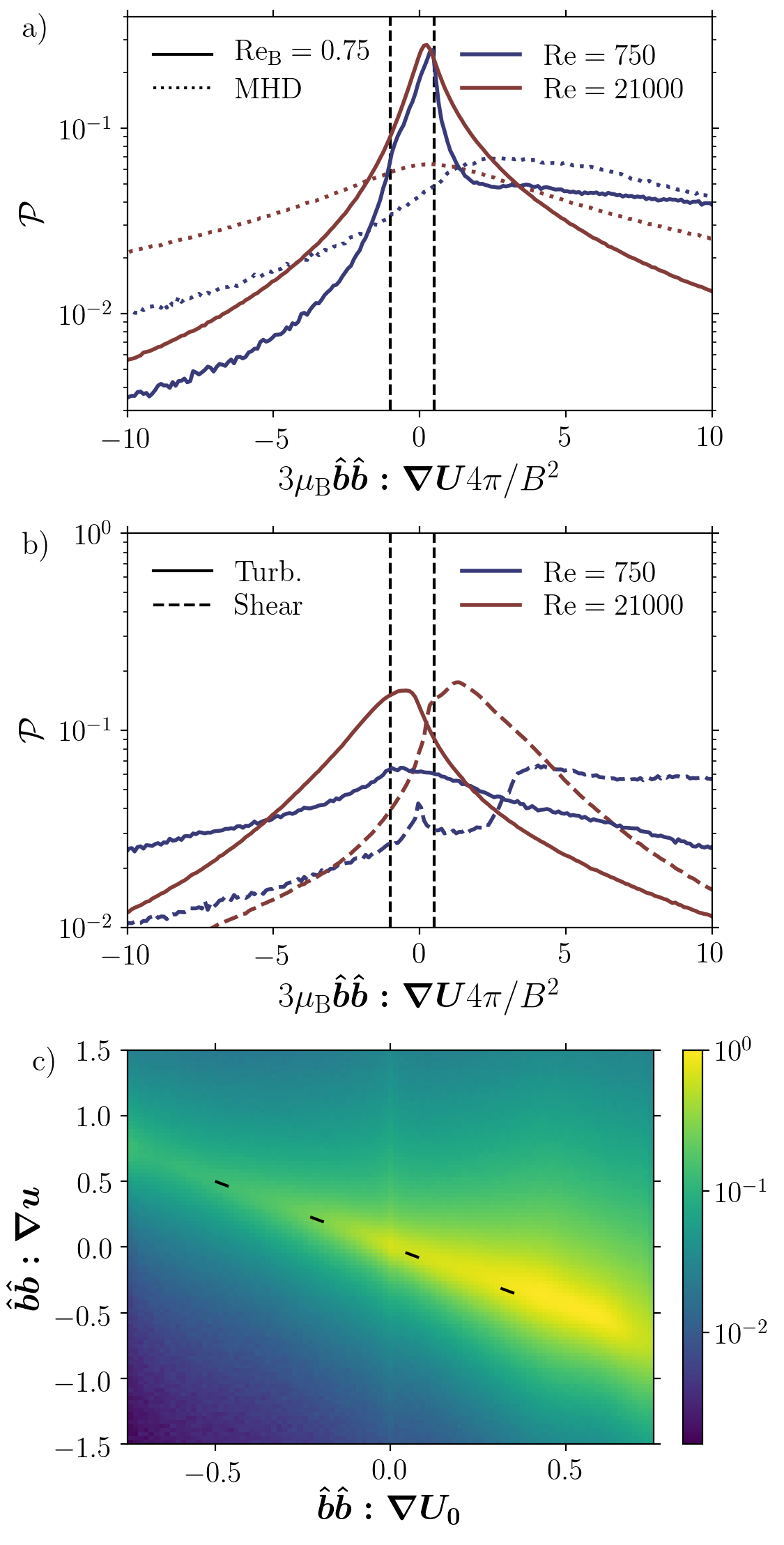}
   \caption{ \textbf{Panel a):}  impact of anisotropic viscosity and isotropic Reynolds numbers on the statistics of the pre-limiter $\dpB$, i.e. $\prel$, in the composite simulations with $\re=750$, $\re=21000$ (both with $\Pm=1$). The dotted lines are MHD at $\orbit = 100$, the solid lines are Braginskii MHD with $\Reb=0.75$ at time $\orbit = 102$. Braginskii viscosity suppresses large $\bbdU$ gradients and drives more cells into the region between the dashed vertical lines (indicating the mirror and firehose limits). The distribution is more symmetric at higher isotropic Reynolds numbers, resulting in a smaller $ \dpm$.    \textbf{Panel b):} Contributions to $\prel$ from the background shear $\bm U_0$ (dashed lines) and the turbulent fluctuations $\bm u$ (solid lines) for $\re = 750$ and $\re = 21000$. Both simulations have $\Pm=1$ and $\Reb = 0.75$. At large $\re$, $\bbdu$ is less constrained by isotropic dissipation, so that anisotropic viscosity more effectively counteracts the positive shear contribution. This produces the more symmetric $\prel$ distribution at large $\re$, shown in the top panel, with suppressed anisotropic transport (Figure \ref{fig:alpha_anis_vs_Re}).  \textbf{Panel c):} 2D histogram of shear and turbulent contributions to $\bbdU$ for the simulation with $\re=21000$, $\Pm=1$ and $\Reb = 0.75$. The dashes indicate the line $\bbdu = - \bbdUs$. The anisotropic viscous stress causes the turbulent $\bbdu$ to balance the largely positive $\bbdUs$ to resist field-line stretching and make the flow magneto-immutable (\citealt{ssqk18}). \label{fig:bbdu_transition} }
\end{figure}

Our composite simulations are summarized in the bottom part of Table \ref{tab:vertical}. 
While we focus on simulations with $\Pm=1$, we also include a simulation with a larger Prandtl number, for which we used $\re=10500$, $\Rem=42000$. The presented values of $\dpmBm$, $\dpBm$ and $\ratio$ are temporal averages, where the averaging is started at time $\orbit = 101$, a time $\Omega^{-1}$ after anisotropic viscosity is added to the system.

We show the evolution of the composite simulations in Figure \ref{fig:transition}, using four different $\re$ with fixed Prandtl number $\Pm=1$. The $\re=750$, 3000, 10500 simulations were performed at resolution $384 \times 192 \times 96$, while for $\re=21000$ we went up in resolution to $768 \times 384 \times 192$. We checked that our $384 \times 192 \times 96$ simulations are converged by also running $\re=10500$ at resolution $576 \times 288 \times 144$. Figure \ref{fig:transition}a shows the evolution of the box-averaged pressure anisotropy; the anisotropic stress evolution is given in Figure \ref{fig:transition}b. The evolution of the anisotropic heating fraction $\Ha$ is shown in Figure \ref{fig:transition}c. The dashed vertical line indicates the time when MHD snapshots were restarted using Braginskii MHD with $\Reb = 0.75$. The MHD $\Dp$, $\Ta$ and $\Ha$ are computed from the MHD flow fields using the same $\nub$ that is used for the Braginskii runs, even though $\Dp$ is not dynamically present in MHD.  

The rapid changes in $\Dp$, $\Ta$, $\Ha$ and subsequent plateau are a convincing demonstration that we reach the Braginskii state very quickly. What is driving the abrupt transition? To understand this, it is instructive to look at $\dpB$ snapshots, before and after anisotropic viscosity is introduced. The upper panels of Figure \ref{fig:snap} show this for the $\re=\Rem=21000$ simulation. In the MHD snapshot in Figure \ref{fig:snap}a, the vast majority of cells are pinned at the mirror/firehose limit. Moreover, there are many ``opposite'' cells in close proximity  of each other. In these regions anisotropic viscosity  operates on a short timescale, trying  to eliminate the strong gradients.  Doing so produces the  smoother distribution of $\Dp$ depicted in Figure \ref{fig:snap}b and causes a rapid change in the volume-averaged $\Dp$. The jump is less pronounced at low $\re$, because the MHD flow field has already been smoothed by isotropic dissipation, thus diminishing the dynamical importance of anisotropic viscosity. In the bottom panels of Figure \ref{fig:snap} we show how in the process of changing the statistics of $\Dp$, Braginskii viscosity also reduces small-scale gradients in the velocity field.

The damping of high-$k$ modes in the Braginskii velocity field leaves an imprint on the power spectrum. We show the kinetic and magnetic energy spectra of the composite simulation with $\re=21000$, $\Pm=1$ in Figure \ref{fig:spectra}, both for the MHD part and the Braginskii MHD ($\Reb =0.75$) part of the run. Figure \ref{fig:spectra}a shows that the kinetic-energy spectra in both models have spectral slopes close to $k^{-3/2}$, the MHD case being slightly shallower. There is also extra suppression of high-$k$ velocity fluctuations in the Braginskii case, which is consistent with the snapshot in Figure \ref{fig:snap}d. Figure \ref{fig:spectra}b demonstrates that the magnetic-energy spectra are hardly modified in Braginskii MHD.

\subsubsection{Dependence on $\re$ and $\Rem$: magneto-immutable  turbulence}

In Figure \ref{fig:bbdu_transition}a we show the pre-limiter anisotropic pressure distribution, before (MHD, dotted line) and after (solid line) the transition. Braginskii viscosity drives more cells into the region inside of the microinstability limits (dashed vertical lines), while damping the tails of the distribution (note that it is the \textit{projection} of $\bm{\nabla u}$ onto the magnetic-field direction, $\bbdu$, that is strongly modified by anisotropic viscosity, $\bb$ and $\bm{\nabla u}$ alone are only mildly affected). In addition, Figure \ref{fig:bbdu_transition}a clearly shows that at low $\re$ we get a distribution that is heavily skewed towards positive $\bbdU$.  At higher $\re$, the distribution becomes more symmetric, leading to a smaller average pressure anisotropy, which explains the results in Figure \ref{fig:transition}.

The results in Figure \ref{fig:transition} and Figure \ref{fig:bbdu_transition}a can be explained qualitatively as follows. At low $\re$, high wavenumber modes of the flow field are efficiently damped by isotropic dissipation and so the fluctuating part of the $\bbdU$ distribution is narrow. Because the shear part of $ \bbdU$ is strongly skewed towards positive values (since $\langle \bbdUs \rangle = \langle -\frac{3}{2}b_{x}b_{y} \Omega  \rangle > 0$), the sum of the two is also going to be biased towards positive values. This leads to a positive and appreciable $\Dp$ approaching the mirror threshold.   

Increasing $\re$ and $\Rem$ means that there are higher wavenumber modes in the turbulent velocity field. In particular, note that $\bm{\nabla u} \sim k^{3/4}$ for a $k^{-3/2}$ spectrum (see Figure \ref{fig:spectra}a), so it increases with increasing resolution. As a result, the fluctuating part of $\bbdU$ becomes more important and can drive more cells towards the firehose limit. Moreover, in Figure \ref{fig:bbdu_transition}b we show  that the $\bbdu$ distribution actually has a negative skew for large-$\re$ turbulence in Braginskii MHD, so that it cancels to a large extent the positive $\bbdUs$ from the mean shear. 

Figure \ref{fig:bbdu_transition}c offers insight into the physics behind the negative skew of the $\bbdu$ distribution. The 2D histogram shows how in the $\re=21000$, $\Reb=0.75$ simulation the plasma rearranges itself to produce a $\bbdu$ that \textit{locally} balances the shear. The fact that the turbulence counters the largely positive $\bbdUs$ can be attributed to anisotropic viscosity causing the rearrangement of the flow field so as to resist changes in $B$ by minimizing $\bbdU$ (\citealt{ssqk18}). We interpret the results of Figure \ref{fig:bbdu_transition} as a consequence of this magneto-immutability: the turbulence can more effectively cancel the field-line stretching of the mean shear ($\bbdUs$) at large $\re$, when the plasma is less constrained by isotropic dissipation. Anisotropic viscosity does also rearrange the turbulence at low $\re$ and $\bbdu$ is able to locally cancel $\bbdUs$ to an appreciable extent (see Figure \ref{fig:bbdu_transition}a for $\re = 750$), but the effect is more pronounced at high $\re$.

\begin{figure*}
\centering
 \includegraphics[scale = 0.5]{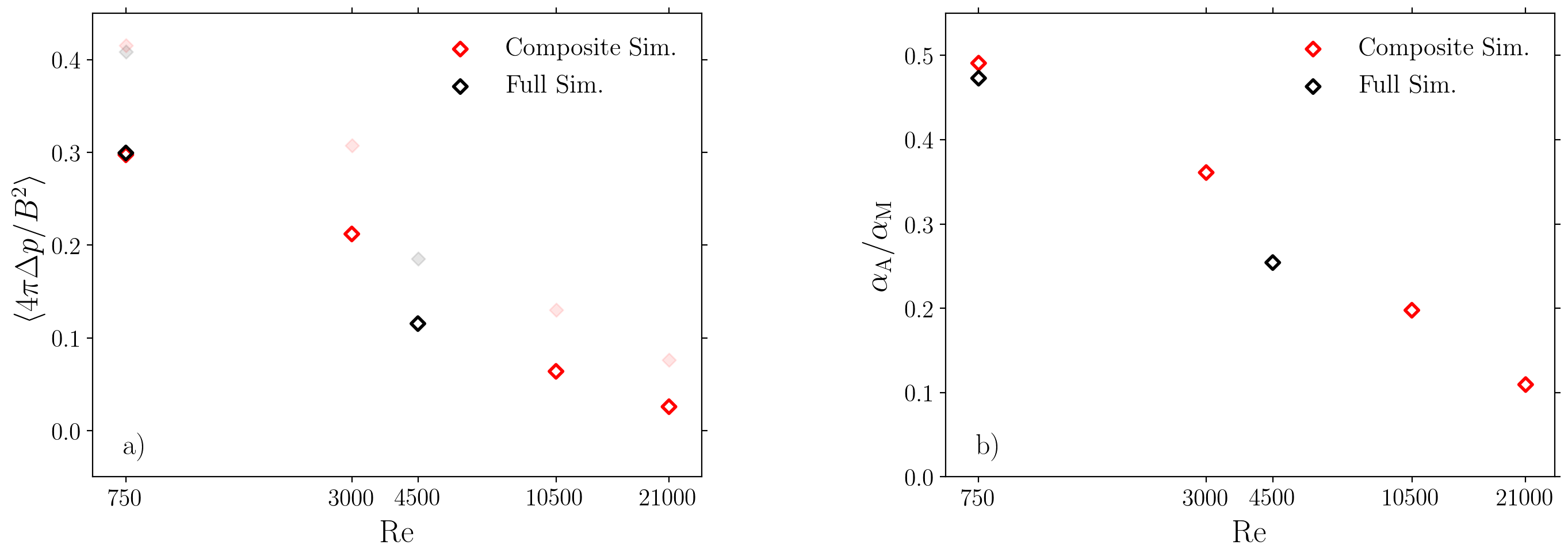}
   \caption{Anisotropic transport and anisotropic pressure vs. $\re$ in simulations with $\Pm=1$ and $\Reb=0.75$. \textbf{Panel a):} $\dpBm$ for different isotropic $\re$. Also shown as filled semi-transparent diamonds are the corresponding values of $\dpmBm$. \textbf{Panel b):} ratio of anisotropic to Maxwell stress vs. $\re$. Composite (red) and full (black) simulations are in good agreement, showing that both anisotropic transport and anisotropic pressure decrease monotonically with isotropic Reynolds number.    \label{fig:alpha_anis_vs_Re} }
\end{figure*}

\begin{figure}
\centering
 \includegraphics[scale = 0.52]{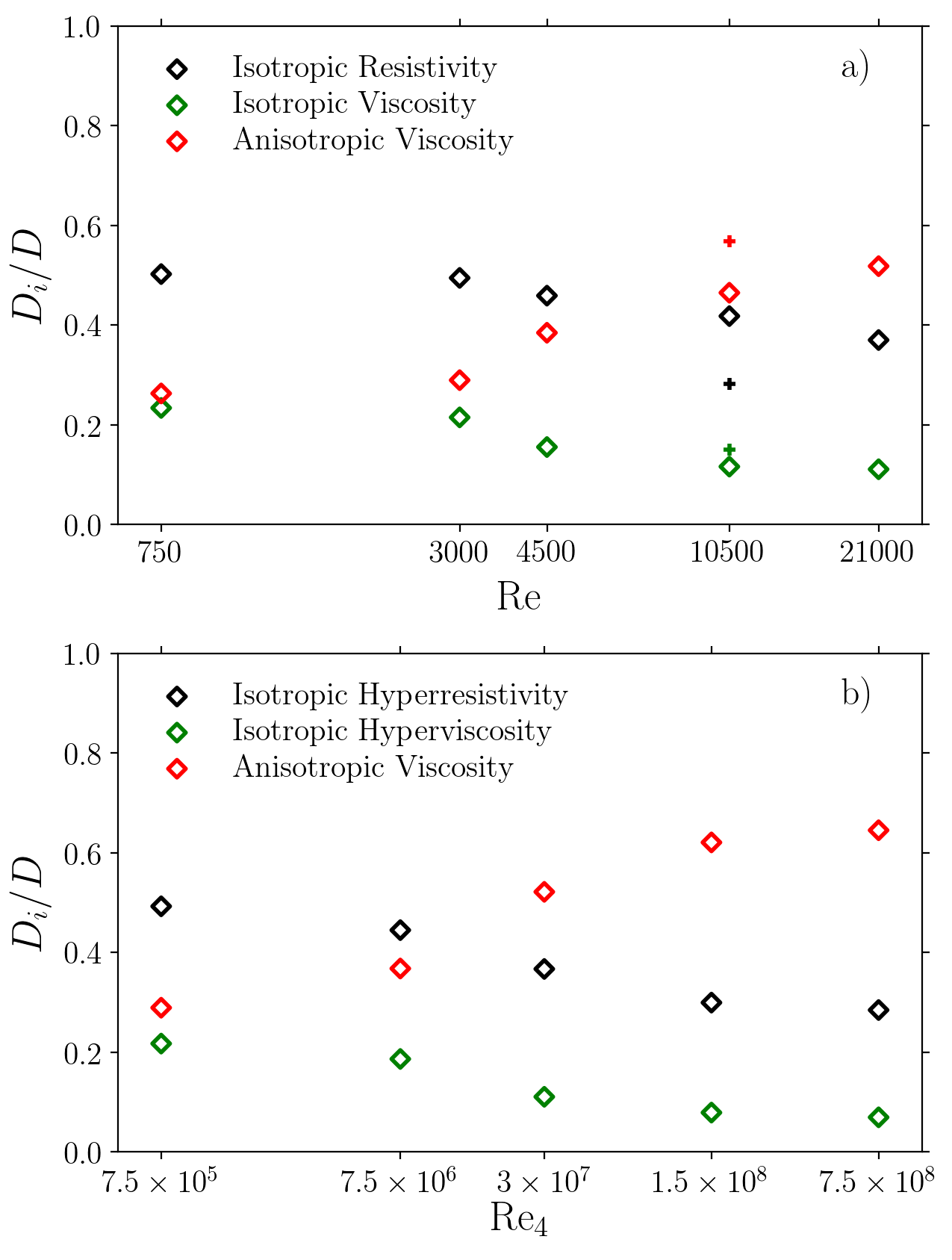}
   \caption{\textbf{Panel a):} temporal averages of heating fractions vs. $\re$ for simulations with $\Reb=0.75$, $\Pm=1$. Heating due to anisotropic viscosity becomes more significant at large isotropic Reynolds numbers, exceeding $50\%$ of the total dissipation. This is also true in our $\re=10500$, $\Pm=4$ simulation, which we show using ``+" markers.  \textbf{Panel b):} Same as the top panel, but for our simulations with hyperdiffusion (with $\Ref = SL_z^4 / \nu_4$), demonstrating that we get qualitatively similar behavior in simulations with hyperdiffusion.    \label{fig:heating_vs_Re} }
\end{figure}

As the  $\bbdU$ distribution becomes broader and more symmetric with increasing $\re$, comparable numbers of cells land on the mirror and firehose sides. As a result, $\dpBm$ and $\ratio$ decrease in value, as in Figure \ref{fig:transition}. $\ratio$  can nevertheless remain more positive in comparison, primarily due to $\langle - B_x B_y \rangle$ typically being larger when averaged over cells at the mirror limit than cells at the firehose limit. It is not entirely surprising that the Maxwell stress is different at the two microinstability boundaries. For example, where the Maxwell stress is negative, the mean shear drives cells towards the firehose side (as $\bbdUs < 0$), whereas $\bbdUs$  is skewed towards the mirror side where $-B_x B_y > 0$. The microinstabilities also significantly affect the dynamical effects of the Maxwell stress: at the firehose limit there is effectively no magnetic tension, and the Maxwell and anisotropic stresses cancel,  while at the mirror boundary the effective magnetic tension is enhanced by a factor $(1 + \dpB)$.   

The values of $\dpBm$ and $\ratio$ as a function of isotropic Reynolds number are plotted in Figure \ref{fig:alpha_anis_vs_Re}a and Figure \ref{fig:alpha_anis_vs_Re}b respectively. In addition to $\dpBm$, Figure \ref{fig:alpha_anis_vs_Re}a also shows the values of $\dpmBm$ as filled semi-transparent diamonds to demonstrate that the exact choice of averaging does not affect our conclusions.  We show all our simulations with $\Pm = 1$, which includes both full Braginskii (black) and composite MHD--Braginskii MHD simulations (red). The two procedures give consistent results in the shared range of $\re$, which supports the plausibility of composite simulation predictions at large Reynolds numbers (see also Appendix \ref{sec:testmixed}).    

Figure \ref{fig:alpha_anis_vs_Re} shows that anisotropic transport and anisotropic pressure decrease significantly as we go to higher $\re$ and $\Rem$\footnote{The trend that $\Ta / \Tm$ and $\dpmBm$ decrease with increasing Reynolds numbers also seems to be present in our two simulations without firehose limiter (see Table \ref{tab:vertical}; note, however, that the two simulations have different $\Pm$). Quite notably, in spite of a large negative pressure anisotropy in the simulation with $\re=10500$ and $\Pm=1$, anisotropic viscous transport remains positive.}. It remains unclear how they will behave as we let $\re$, $\Rem$ $\rightarrow \infty$. Unfortunately probing larger isotropic Reynolds numbers is beyond our current computational capabilities.


In spite of $\dpmBm$, $\dpBm \rightarrow 0$ at large $\re$, anisotropic pressure is an important source of dissipation. This is because regions at the firehose and mirror boundaries contribute positively to the  anisotropic viscous heating  rate (eq. \ref{eq:anis_diss}), as $\Dp$ and $\bbdU$ have the same sign (see eq. \ref{eq:delp}). In Figure \ref{fig:heating_vs_Re}a we show that anisotropic viscosity becomes the dominant source of dissipation at large $\re$, even though the volume-averaged pressure anisotropy is a steadily decreasing function of $\re$ (Figure \ref{fig:alpha_anis_vs_Re}).

\subsubsection{Simulations with hyperdiffusion} 
\label{sssec:vs_Ref}

\begin{figure*}
\centering
 \includegraphics[scale = 0.5]{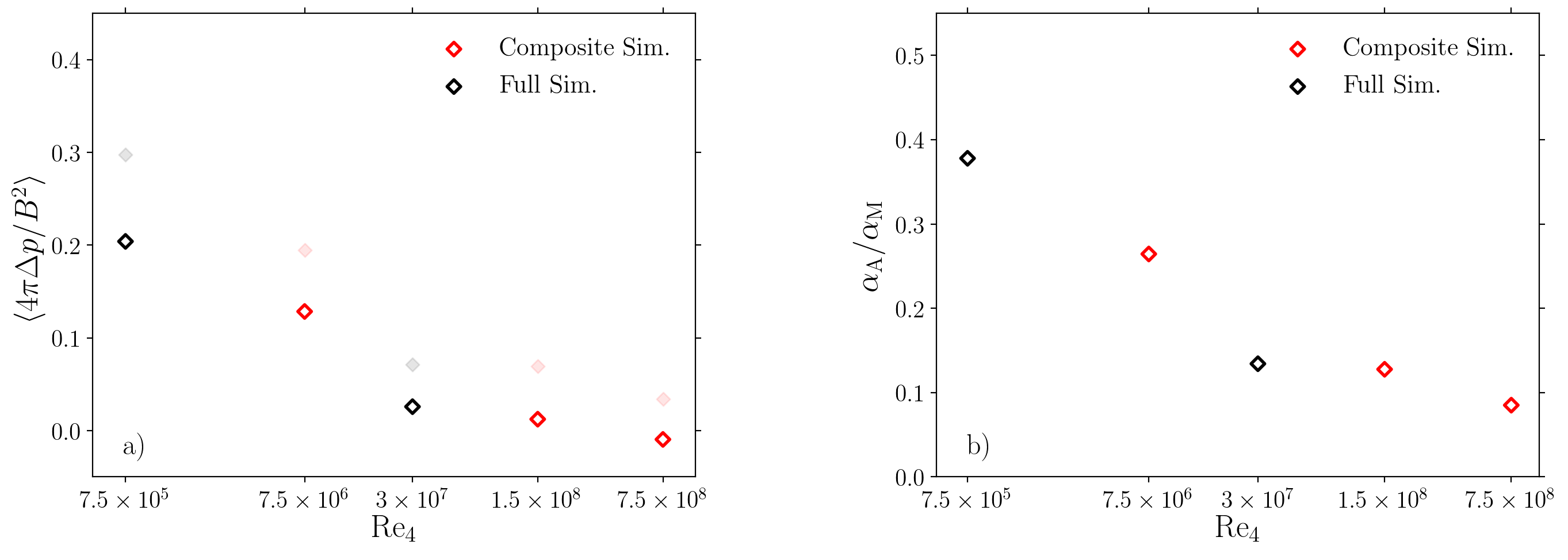}
   \caption{Same as Figure \ref{fig:alpha_anis_vs_Re} but with fourth-order hyperdiffusion. As before, we show the values of $\dpmBm$ in the left panel as filled semi-transparent diamonds. At small $\Ref = S L_z^2 / \nu_4$ we get a dependence qualitatively similar to second-order diffusion. At large $\Ref$ we obtain a plateau-like shape with $\dpBm \sim 0$, $\dpmBm \sim 0$ and $\ratio \sim 0.1$.    \label{fig:alpha_anis_vs_nu4} }
\end{figure*}

By increasing the range of scales over which viscosity and resistivity are negligible, simulations with isotropic diffusion replaced by fourth-order hyperviscosity ($\nu_4 \nabla^4 \bm{u}$) and hyperresistivity ($\eta_4 \nabla^4  \bm{B}$) provide an alternative, indirect way of probing larger effective isotropic Reynolds numbers that does not require higher resolutions. We therefore augment our $\Pm=1$ simulations with hyperdiffusion simulations with varying  
\begin{equation} \label{eq:ref}
\Ref= SL_z^4/\nu_4.
\end{equation}
All simulations have a hyper-Prandtl number equal to 1, i.e. $\nu_4 = \eta_4$, and we summarize them in Table \ref{tab:hyper}.

We show the $\dpBm$ and $\ratio$ of our simulations with hyperdiffusion in Figure \ref{fig:alpha_anis_vs_nu4}.  At small $\Ref$, $\dpBm$ and  $\ratio$ follow a similar trend to that shown in Figure \ref{fig:alpha_anis_vs_Re}. For large $\Ref$, we obtain a plateau-like shape, with $\dpBm \sim 0$ and $\ratio \sim 0.1$. The plateau is consistent with the $\re=21000$, $\Pm=1$ simulation, which suggests that this may already be close to the asymptotic limit when $\re$, $\Rem$ $\rightarrow \infty$.

\begin{table*}
\centering
\caption{Summary of Braginskii MHD simulations with net vertical magnetic field and fourth-order isotropic hyperdiffusion. \textbf{Full} simulations were evolved for a time $\orbit = 100$ and the values of $\dpmBm$, $\dpBm$ and $\ratio$ are temporal averages over  $\orbit = 70-100$. In \textbf{Composite} simulations, MHD fields are restarted at $\orbit= 100$ using Braginskii MHD. $\dpmBm$, $\dpBm$ and $\ratio$  were averaged over $\orbit = 101-110$, except for the $\Ref = 7.5 \times 10^8$ simulation, for which the average is over $\orbit = 101-102$.}
\label{tab:hyper}  
\begin{tabular}{ccccccccc}\hline\hline
Resolution & Sim. Type & $\Ref$ & $\Remf$ & $\nu_4 / \eta_4$ & $\Reb$ & $\dpmBm$  & $\dpBm$ & $\ratio$ \\
\hline\hline
\\
$(256,128,64)$ &Full & $7.5 \times 10^5$  & $7.5 \times 10^5$  & 1 & 0.75 &  0.30  & 0.20 & 0.38   \\
$(256,128,64)$ &Composite & $7.5 \times 10^6$  & $7.5 \times 10^6$  & 1 & 0.75 & 0.19  & 0.13 & 0.26 \\
$(256,128,64)$ &Full & $3 \times 10^7$  & $3 \times 10^7$  & 1 & 0.75 &  0.071   &0.026 &  0.13   \\
$(384,192,96)$ &Composite & $1.5 \times 10^8$  & $1.5 \times 10^8$ & 1 & 0.75 & 0.069   &0.012 &  0.13   \\
$(576,288,144)$ &Composite & $7.5 \times 10^8$ & $7.5 \times 10^8$ & 1 & 0.75 & 0.034  & -0.0096 & 0.085 \\
\\
\hline
\hline
\end{tabular}
\end{table*}

In Figure \ref{fig:heating_vs_Re}b we show that anisotropic viscous heating also increases with decreasing isotropic hyperdiffusivities, accounting for more than $60\%$ of the total dissipation in our largest-$\Ref$ simulations (hyperviscous and hyperresistive heating are calculated as $ \nu_4 \Cal \bm{u \cdot } \nabla^4 \bm{u }$ and $ \eta_4 / (4\pi) \Cal \bm{B \cdot } \nabla^4 \bm{B }$ respectively). 

We end this section by pointing the reader to Appendix \ref{sec:mod_hyp4}, where we discuss significant increases in $\Tm$ and $\Tr$ that happen soon after Braginskii viscosity is introduced in the simulations with $\Ref = 1.5 \times 10^8$ and  $\Ref = 7.5 \times 10^8$. While the origin of the enhanced transport remains unknown, we suspect that it is a numerical effect that is specific to hyperdiffusion operators (possibly overemphasis of channel modes in the presence of Braginskii viscosity) because our high-$\re$ second-order-diffusion simulations did not give any indications of similar behavior.

\section{ Other Field Configurations}
\label{sec:otherfield}

\subsection{Net Toroidal and Vertical Field}
\label{sec:toroid}

\begin{figure}
\centering
 \includegraphics[scale=0.5]{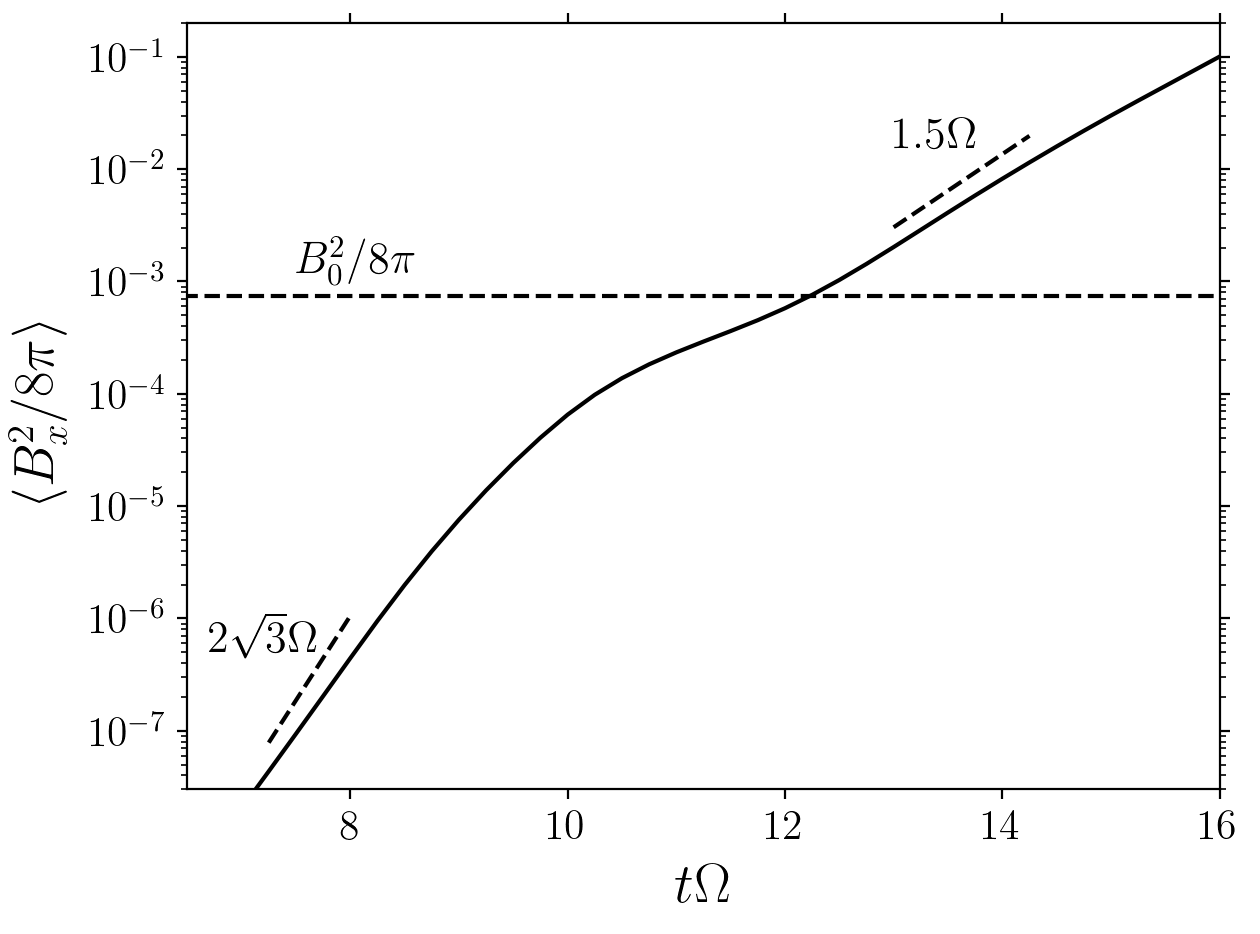}
   \caption{Linear and nonlinear evolution of the x-component of magnetic energy in Braginskii MHD with $\Reb=0.75$ for $\langle B_{y} \rangle = \langle B_{z} \rangle$. In the linear phase the  growth rate exceeds the MHD growth rate and agrees with the analytic prediction. In the nonlinear phase, when the pressure-anisotropy limiters become important, it reverts to the MHD growth rate.    \label{fig:growth_transition} }
\end{figure}

\begin{figure}
\centering
 \includegraphics[scale=0.57]{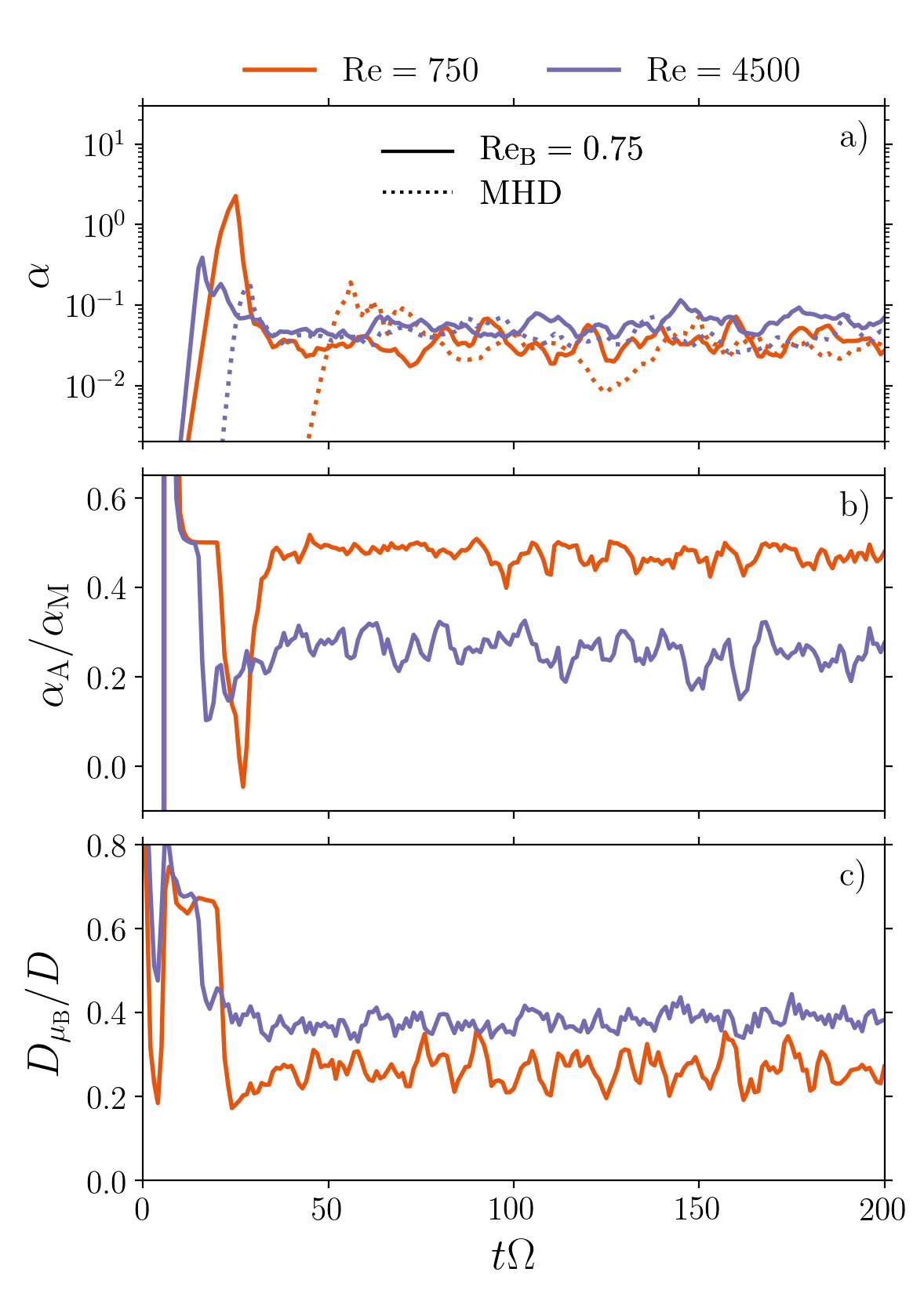}
   \caption{Evolution in a box  with $\langle B_y \rangle = \langle B_z \rangle$ for $\re=750$ and $\re=4500$ (both with $\Pm=1$). \textbf{Panel a):} evolution of $\alpha$ in Braginskii MHD with $\Reb=0.75$ (solid) and MHD (dotted). \textbf{Panel b):} evolution of $\ratio$ in Braginskii MHD for the two $\re$, again demonstrating that anisotropic transport is less significant at large Reynolds numbers. \textbf{Panel c):} evolution of the anisotropic viscous heating fraction in the two simulations. $\Ha$ increases with increasing $\re$, which is consistent with our results for net vertical flux alone in Section \ref{sec:vertical} (Figure \ref{fig:heating_vs_Re}).    \label{fig:toro_full} }
\end{figure}

In the previous section we have seen that transport properties in MHD and Braginskii MHD are remarkably similar, at least for the case of net vertical flux. Transport is modified primarily by the addition of anisotropic stress, but its importance decreases at large isotropic $\re$ and $\Rem$.  Anisotropic viscosity does significantly influence the heating of the plasma and the statistics of $\bbdU$,  but it does not notably alter the turbulent amplitudes and transport properties of the flow. 

In this section we look at a mixed vertical-azimuthal field geometry, which exhibits different linear behavior in Braginskii MHD and MHD, and so is arguably the most likely to show significant differences in nonlinear transport properties. More specifically, we use the field configuration,

\begin{equation}
\langle \bm{B} \rangle = B_0 (\ey + \ez ),
\end{equation}
where $B_0$ is the same as in Section $\ref{sec:vertical}$. In this field configuration with equal toroidal and vertical components, \cite{qdh02} and  \cite{b04} have shown that anisotropic pressure increases the linear growth rate of the MRI in Keplerian discs from $0.75 \Omega$ to $\sqrt{3} \Omega$. Here we try to address whether there are any differences in the nonlinear dynamics.

We run our vertical-azimuthal-field simulations in boxes of size $L_x = 4$, $L_y = 4$ and $L_z = 1$, just like for the net-vertical-flux case described in Section \ref{sec:vertical}. While in this field configuration the MRI modes are larger in scale, we have found that this box size is sufficient to capture the fastest-growing MRI wavelengths (Figure \ref{fig:growth_transition} supports this) and the fastest-growing parasitic modes (similar to the net-vertical-field case, horizontally narrow $1\times4 \times 1$ boxes did not correctly capture the most important parasitic modes).   

First we focus on how the MRI enters the nonlinear growth phase. Figure \ref{fig:growth_transition} shows the evolution of the MRI in Braginskii MHD with $\Reb=0.75$, as it transitions from linear growth to the nonlinear regime. In order for the growth rates to be close to the ideal values, we are using hyperdiffusion to dissipate energy just above the grid scale. The initial growth rate of $\langle B_x^2 \rangle$ is close to the theoretical value of $2\sqrt{3} \Omega$. Once the perturbed magnetic energy becomes comparable to the energy in the background field, the presence of anisotropic-pressure limiters becomes significant. This limits the dynamical importance of the anisotropic stress, which causes the growth rate to decrease to its MHD value and the fastest-growing mode to migrate to shorter wavelengths. See \cite{sqk17b} for more discussion.

In Figure \ref{fig:toro_full} we show the complete evolution into turbulence for $\re=750$ and $\re=4500$, both with $\Pm=1$. The evolution of $\alpha$ is shown in Figure \ref{fig:toro_full}a. The dotted line represents the MHD solution and the solid line is the $\Reb=0.75$ evolution. Due to the faster linear growth phase, Braginskii MHD saturates into turbulence earlier, but the final turbulent state is again similar to the MHD solution. 

We show the evolution of $\ratio$ in Figure \ref{fig:toro_full}b. We have chosen the same $\re$ as in the full Braginskii simulations with net vertical field (see Figure \ref{fig:alpha_anis_vs_Re}). The result is qualitatively the same: $\ratio$ is smaller at large isotropic Reynolds numbers. Heating due to anisotropic viscosity (Figure \ref{fig:toro_full}c) nonetheless increases with increasing $\re$, which is consistent with our simulations with net vertical field.  The temporal averages of $\ratio$ and $\Ha$ obtained here are also very similar to the ones in Figure $\ref{fig:alpha_anis_vs_Re}$ and Table \ref{tab:vertical}.       

In addition to our $\Pm=1$ simulations, we have also evolved the $\langle B_y \rangle = \langle B_z \rangle$ field for other choices of Prandtl number. For temporal averages of transport coefficients we refer the reader to Table \ref{tab:toro}. 

 Just like for the case of net vertical flux, we find that the total level of angular-momentum transport is not greatly affected by the presence of anisotropic pressure. The decrease in anisotropic transport at high Reynolds numbers, accompanied by increased anisotropic viscous heating, also persists in boxes threaded by equal vertical and azimuthal fields.

\begin{table*}
\centering
\caption{Summary of simulations with net toroidal and vertical magnetic field ($\langle B_y \rangle = \langle B_z \rangle$). The transport coefficients and mean pressure anisotropies were averaged over $\orbit = 100 - 200$.  }
\label{tab:toro} 
\begin{tabular}{ccccccccccccc}\hline\hline
Resolution & $\re$ & $\Rem$ & $\Pm$ & $\Reb$ & $\Tr$ & $\Tm$ & $\Ta$ & $\alpha$ & $\dpmBm$  & $\dpBm$ & $\ratio$   \\
\hline\hline
\\
$(256,128,64)$ & $3000$ & 1500 & 0.5 & -- & 0.0056   & 0.022   &  -- & 0.028 & -- & -- & --  \\
$(256,128,64)$ & $3000$ & 1500 & 0.5 & 0.75 & 0.0064   & 0.019   & 0.0076  &    0.033 & 0.32 &  0.24 & 0.40\\
\\
$(256,128,64)$ & $750$ & 750 & 1 & -- & 0.0059    & 0.023   &  -- & 0.029 & -- & -- & --    \\
$(256,128,64)$ & $750$ & 750 & 1 & 0.75 & 0.0064   & 0.020    & 0.0094  & 0.036 & 0.41 & 0.30   & 0.47  \\
\\
$(256,128,64)$ & $4500$ & 4500 & 1 & -- & 0.0067   & 0.033    &  -- & 0.040  & -- & -- & --   \\
$(256,128,64)$ & $4500$ & 4500 & 1 & 0.75 & 0.011    & 0.042   & 0.011  & 0.063 & 0.18  & 0.11 & 0.25   \\
\\
\hline
\\
$(192,96,48)$ & $750$ & 3000 & 4 & -- & 0.011  & 0.067 & -- & 0.078 & -- &-- & --   \\
$(192,96,48)$ & $750$ & 3000 & 4 & 0.3 & 0.014 & 0.063  & 0.022 & 0.099  & 0.28 & 0.19 & 0.36  \\
\\
\hline
\hline
\end{tabular}
\end{table*}

\subsection{Zero Net Flux}
\label{sec:zero}
In Section \ref{sec:vertical} we showed that the Prandtl-number dependence of ordinary MHD is recovered in Braginskii MHD with net vertical field.
We now turn our attention to the zero-net-flux case, where there is no mean magnetic field threading the box. As is common in previous literature, we start from the initial condition 
\begin{equation}
\bm{B}(t=0) = B_0 \sin\Big(\frac{2 \pi x}{L_x}\Big) \ez .
\end{equation}
\cite{fplh07} showed that whether turbulence is sustained or decaying in zero-net-flux simulations is highly sensitive to the isotropic Reynolds and Prandtl numbers. Given the viscous nature of the pressure anisotropy, it is instructive to see if this behavior is modified in Braginskii MHD. More specifically, if anisotropic pressure were to modify the effective Prandtl number, we should see a clear imprint near the turbulence/no turbulence boundary.

\begin{table*}
\centering
\caption{Summary of simulations with zero net flux.  The transport coefficients and the mean pressure anisotropies were averaged over $\orbit = 100 - 200$. The last column indicates whether turbulence could be sustained (``SD'' means that the turbulence is slowly decaying). See Figure \ref{fig:zeronet_evol} for examples. \label{tab:zeronet}}

\begin{adjustbox}{width=1.0\textwidth}

\begin{tabular}{ccccccccccccc}\hline\hline
Box Size & Resolution & $\re$ & $\Rem$ & $\Pm$ & $\Reb$ & $\Tr$ & $\Tm$ & $\Ta$ & $\alpha$  & $\dpBm$ & $\ratio$ & Turb? \\
\hline\hline
\\
$(4,4,1)$&$(256,128,64)$ & $1500$ & 12000 & 8 & -- & 0.0020   & 0.015  & -- &0.017  & -- & -- &  Yes  \\
$(4,4,1)$&$(256,128,64)$ & $1500$ & 12000 & 8 & 1.5 & 0.0040 & 0.020 &0.0078  & 0.032   & 0.17 &0.39 & Yes\\
\\
$(4,4,1)$&$(256,128,64)$ & $6000$ & 12000 & 2 & -- &   &   &  -- &   & -- & -- & SD  \\
$(4,4,1)$&$(256,128,64)$ & $6000$ & 12000 & 2 & 1.5 &   &   &  & &   & & SD  \\
\hline
\\
$(1,4,1)$&$(64,128,64)$ & $750$ & 6000 & 8  & -- &   &   &  -- &   & -- & -- & No  \\
$(1,4,1)$&$(64,128,64)$ & $750$ & 6000 & 8 & 0.75 &   &   &  &   & & & No  \\
\\
$(1,4,1)$&$(64,128,64)$ & $1500$ & 6000 & 4  & -- &   &   &  -- &   & -- & -- &  No  \\
$(1,4,1)$&$(64,128,64)$ & $1500$ & 6000 & 4 & 0.75 &   &   &  & &    & & No \\
$(1,4,1)$&$(64,128,64)$ & $1500$ & 12000 & 8  & -- & 0.0012   & 0.0091  &  -- & 0.010   &-- &-- & Yes   \\
$(1,4,1)$&$(64,128,64)$ & $1500$ & 12000 & 8 & 0.75 & 0.0021   & 0.011  & 0.0047  & 0.017  & 0.21 & 0.45 & Yes  \\
\\
$(1,4,1)$&$(64,128,64)$ & $3000$ & 6000 & 2  & -- &   &   &  -- &    &  -- & -- & No   \\
$(1,4,1)$&$(64,128,64)$ & $3000$ & 6000 & 2 & 0.75 &   &   &  &  &   &  & No  \\
$(1,4,1)$&$(64,128,64)$ & $3000$ & 12000 & 4 & -- & 0.0008  & 0.0054   &  -- & 0.0061   & -- & -- & Yes  \\
$(1,4,1)$&$(64,128,64)$ & $3000$ & 12000 & 4 & 0.75 &0.0002   & 0.0010   & 0.0005  &  0.0017 & 0.24 & 0.53 & Yes  \\
\\
\hline
\hline
\end{tabular}
\end{adjustbox}
\end{table*}

Most of our zero-net-flux simulations are in boxes of dimensions $1 \times 4 \times 1$, which is similar to \cite{fplh07}. We also ran a few simulations in horizontally extended boxes ($4 \times 4 \times 1$), although this does not significantly modify the bulk transport properties in the case of of zero net flux (unlike magnetic-field geometries with a net vertical component). In both cases $B_0$ is chosen such that the initial average magnetic energy is the same as in the net-vertical-flux case described in Section \ref{sec:vertical}. 

Table \ref{tab:zeronet} gives a summary of our zero-net-flux simulations. Once again, we do not find any systematic new behavior introduced by anisotropic viscosity. The time-averaged values of $\alpha$  may differ by order unity, but there does not seem to be a general trend.

\begin{figure}
\centering
 \includegraphics[scale = 0.35]{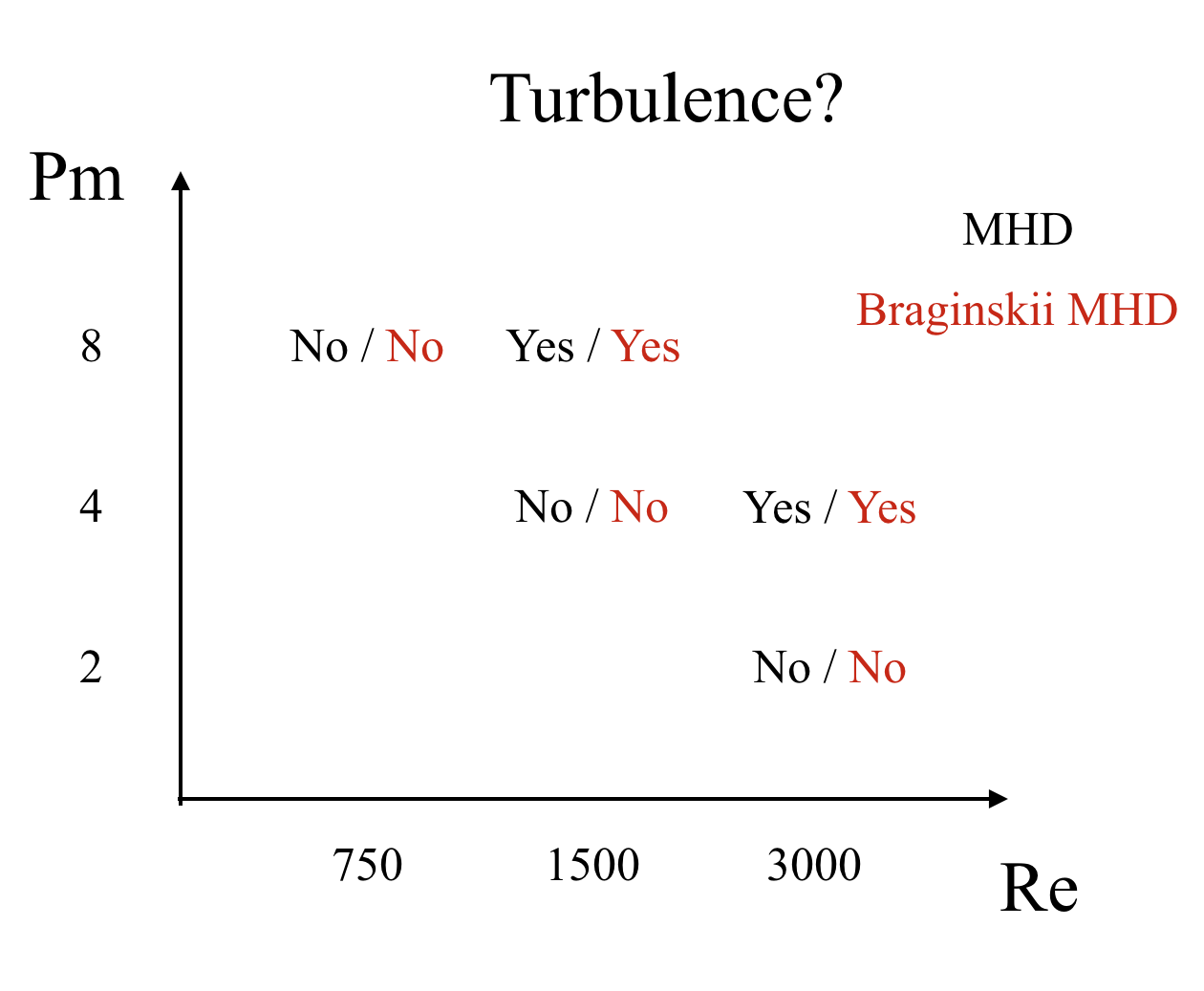}
   \caption{Zero-net-flux turbulence/no turbulence boundary in the $\re-\Pm$ plane in MHD and Braginskii MHD with $\Reb=0.75$. ``No'' means that turbulence eventually decayed; ``Yes'' indicates a $\re$, $\Pm$ pair where turbulence could be sustained. See Figure \ref{fig:zeronet_evol} for examples.    \label{fig:turb_boundary} }
\end{figure}

In Figure \ref{fig:turb_boundary} we show the turbulence/no turbulence boundary in $\re-\Pm$ space, in MHD and in Braginskii MHD with $\Reb = 0.75$. This can be compared to Fig. 11 in  \cite{fplh07}, who show the analogous boundary for compressible MHD. In spite of the additional large anisotropic viscosity, the turbulence boundary is apparently unchanged in Braginskii MHD.   

\begin{figure}
\centering
 \includegraphics[scale = 0.5]{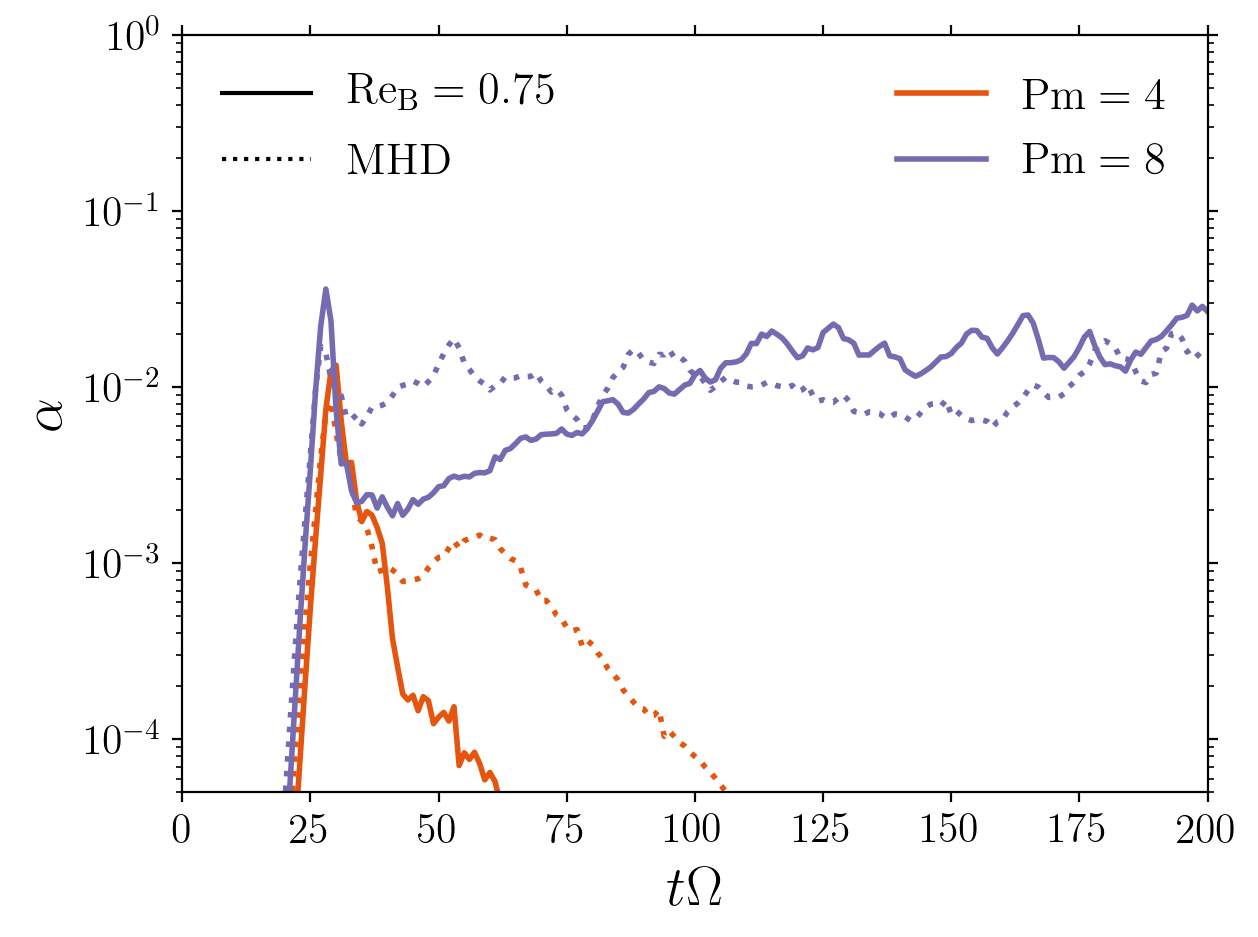}
   \caption{Evolution of $\alpha$ in a box with zero net flux, in MHD (dotted) and Braginskii MHD with $\Reb = 0.75$ (solid). We show the evolution for $\re=1500$ with two different isotropic Prandtl numbers: $\Pm=4$ and $\Pm=8$. Like in MHD, low-Prandtl-number turbulence decays in Braginskii MHD and only high-Prandtl-number turbulence can be sustained.    \label{fig:zeronet_evol} }
\end{figure}

Figure \ref{fig:zeronet_evol} shows the evolution of angular-momentum transport in some of our zero-net-flux simulations. While the low-Prandtl-number simulation decays both in MHD and in Braginskii MHD, at high Prandtl number both models can sustain turbulence.  The turbulence/no turbulence behavior does not change when Braginskii viscosity is present. 

\section{Conclusions}
In this paper we have examined the importance of anisotropic pressure for MRI-driven turbulence in a shearing box. This was achieved through a combination of full incompressible Braginskii MHD simulations and composite MHD--Braginskii MHD simulations, in which fully turbulent MHD flow fields were restarted with Braginskii viscosity included, in order to minimize the computational cost due to the large viscosity (Appendix \ref{sec:testmixed} validates this method). We find that bulk transport properties of MRI turbulence are effectively unchanged by large Braginskii viscosities. This is at first glance surprising given that the pressure-anisotropy stresses are at least as important as other forces in the system and modify the flow structures in the turbulence.

We looked at three initial magnetic-field orientations, including net vertical flux, equal net toroidal and vertical field, and zero net flux. All were examined using a number of isotropic Reynolds and magnetic Reynolds numbers. In our simulations, we augmented the Braginskii equations (\ref{eq:continuity}--\ref{eq:delp}) to account for the presence of kinetic microinstabilities: the mirror and firehose instabilities, which are excited when the pressure anisotropy becomes comparable to the magnetic pressure (equations \ref{eq:mirror} \& \ref{eq:firehose}). As these instabilities tend to pin $\Dp$ near marginal stability (\citealt{ksq16}), we include this kinetic result by introducing hard-wall limits on $\Dp$ in our fluid simulations at the instability thresholds.  

We can divide our results into two main categories. Our first result concerns the level of angular-momentum transport in Braginskii MHD. We find that, for the range of isotropic diffusivities tested in this work, the Maxwell and Reynolds components of the stress tensor are hardly modified in Braginskii MHD. The presence of Braginskii viscosity modifies transport primarily through its additional anisotropic viscous stress component. This anisotropic viscous stress, however, is consistently smaller than the Maxwell stress, so that the total angular-momentum transport is only mildly affected. 

One interesting consequence of this is that in Braginskii MHD transport remains very sensitive to isotropic diffusivities, despite the fact that the anisotropic viscosity is orders of magnitude larger than the isotropic viscosity. For the range of Reynolds numbers probed, the Maxwell and Reynolds stresses exhibit the same magnetic-Prandtl-number dependence  as in ordinary MHD in all of the tested magnetic-field configurations. This includes the temporal averages of $\alpha$ in net-flux simulations (see Figure \ref{fig:alphaPm}), as well as the turbulence/no turbulence boundary in the $\re-\Pm$ plane for zero-net-flux simulations (Figure \ref{fig:turb_boundary}). Remarkably, the angular-momentum transport in the shearing box, which is sensitive to a number of parameters (e.g. $\Pm$, box size), appears to be almost unaffected by even large Braginskii viscosities. 

Our systematic demonstration of MHD-like behavior in Braginskii MHD complements and clarifies some of the findings made by \cite{shqs06}, \cite{ksq16}, \cite{sqk17b} and \cite{fcgqt17}, who have seen strong similarities to MHD in their extended-MHD and kinetic simulations. Nevertheless, we do find that anisotropic viscosity leaves a clear imprint on the structure of the flow. $\Dp$ is an important source of dissipation (Figure \ref{fig:heating_vs_Re}) and strongly modifies the velocity component along the local magnetic field  (Figure \ref{fig:bbdu_transition}). Surprisingly, this turns out to not significantly affect angular-momentum transport.   

Our second primary result is related to the anisotropic viscous stress component. We have demonstrated that anisotropic transport becomes independent of anisotropic viscosity at sufficiently large anisotropic viscosity $\nub$ (see Figure \ref{fig:alpha_nuB}), but is still sensitive to the amount of isotropic dissipation. This dependence is rather significant. We recover the typical result found in previous works -- that $\Ta \sim \Tm$ -- \textit{only at low isotropic Reynolds numbers}. We find that the ratio $\ratio$ decreases systematically with $\re$ and $\Rem$, down to $\sim 0.1$ for $\re=\Rem=21000$. This is driven primarily by a decreasing average pressure anisotropy, as shown in Figure \ref{fig:alpha_anis_vs_Re}. Simulations with viscosity and resistivity replaced by hyperdiffusion operators suggest that $\ratio \sim 0.1$ and $\dpmBm$, $\dpBm \sim 0$ may be close to the asymptotic limit as $\re$, $\Rem \rightarrow \infty$ (see Figure \ref{fig:alpha_anis_vs_nu4}).  This has not been seen in any previous studies, which have consistently found an anisotropic stress that is comparable to the Maxwell stress (\citealt{shqs06}; \citealt{ksq16}; \citealt{fcgqt17}). However, previous fluid simulations with anisotropic pressure were at lower resolution and were grid-based rather than spectral, so it is likely that they correspond to our low-$\re$ simulations. Although high resolutions were used in  the fully kinetic simulations of \cite{ksq16}, it is unclear whether the different $\ratio$ seen in this case arises from differences between weakly collisional and collisionless plasmas, or the difficulty of maintaining large scale separations between the large-scale MRI-generated motions and plasma microinstabilities. 

Even though $\dpBm$ decreases with increasing $\re$, anisotropic pressure becomes a more important source of dissipation at large isotropic Reynolds numbers, accounting for more than $50\%$ of the total heating at our highest $\re$. We show this in Figure \ref{fig:heating_vs_Re}. 

We interpret the decreasing $\dpBm $ and $\ratio$ with increasing $\re$ as being due to the influence of isotropic dissipation on the statistics of $\bbdU$, thus affecting the box-averaged pressure anisotropy $\dpm$ via equation \eqref{eq:delp}. Both the background shear $\bm{U_0}$ and the perturbed turbulent velocity field $\bm{u}$ contribute to $\bbdU$. The background shear contribution, $\bbdUs = -\frac{3}{2}b_x b_y \Omega$, is largely positive. Meanwhile, $\bbdu$ is sensitive to the choice of Reynolds numbers. At low $\re$ with more isotropic damping, the distribution of $\bbdu$ across the simulation domain is narrow, so that the mean shear leads to a skewed $\bbdU$ distribution with the vast majority of cells on the mirror side. This produces a large $\dpm$ close to the mirror threshold and $\Ta \sim \Tm$. In simulations with higher $\re$, $\bbdu$ is broader and more important, as higher wavenumber modes are present in the turbulent velocity field and $\bm{\nabla u} \sim k^{3/4}$  for a $k^{-3/2}$ spectral slope (see Figure \ref{fig:spectra}a). The width of $\bbdu$ increases with increasing $\re$ despite the large $\nub$, as  the damping of  high-$k$ modes in $\bbdu$ is significantly inhibited by the anisotropic-pressure limiters (since the highest-$k$ modes typically correspond to unlimited $\Dp \gg B^2 / 4\pi$, see Figure \ref{fig:bbdu_transition}a). As a result, the fraction of cells with a negative $\Dp$ increases with increasing $\re$, which reduces the box-averaged pressure anisotropy. In addition, because the plasma is less constrained by isotropic dissipation at large $\re$, anisotropic viscosity can also more effectively reorganize the turbulence to balance the positive $\bbdUs$ (shown in Figure \ref{fig:bbdu_transition}c). We interpret this as being related to the idea of magneto-immutability described in \cite{ssqk18}, who show that anisotropic viscosity acts to minimize field-line stretching $\bbdU$ to resist changes in magnetic-field strength. For these reasons, high $\re$ turbulence produces a more symmetric $3\nub \bbdU /B^2$ distribution centered closer to zero, and with comparable number of cells around the mirror and firehose limits (driven by the mean shear and turbulent fluctuations, respectively). This causes $\dpm$ and $\Ta$ to decrease with increasing $\re$. We illustrate this behavior in Figure \ref{fig:bbdu_transition}, showing the distributions of $\prel$ for small and large $\re$. 

Although $\dpBm \rightarrow 0$ at large $\re$, the fractional heating due to anisotropic viscosity increases at large $\re$ (Figure \ref{fig:heating_vs_Re}). This is because there is positive heating at both the firehose and mirror boundaries (since $\Dp$ and $\bbdU$ in eq. \ref{eq:anis_diss} have the same sign). Note also that $\langle |\Dp| \rangle \sim \langle B^2 / 4\pi \rangle$ and $\langle | \bbdU | \rangle$ both increase with increasing $\re$. 

While our simulations offered some insight, we must nonetheless still speculate about what happens when $\re, \Rem \rightarrow \infty$.  Can $\Ta$ and $\dpm$ become negative? Or do they perhaps both tend to $0$? Both cases would have interesting consequences, the latter implying that $\alpha_{\mathrm{Braginskii}} \approx \Tr + \Tm \rightarrow \alpha_{\mathrm{MHD}}$, provided the Braginskii Maxwell and Reynolds stresses continue to track the MHD result  (which is not entirely certain, see e.g. Appendix \ref{sec:mod_hyp4}). Our simulations with hyperdiffusion tentatively suggest that $\dpmBm$, $\dpBm \sim 0$ and $ \ratio  \sim 0 \leftrightarrow 0.1$ are potential candidates for the asymptotic values. Nevertheless, exploring the asymptotic limit of large Reynolds numbers in detail is beyond the scope of this paper and a potential direction for future studies. This, however, would most likely require a more efficient Braginskii viscosity implementation than  the one used in this work. Another interesting extension to this work would be to examine the compressible Braginskii MHD equations, including the effects of anisotropic conduction.

It is unclear if and how our results apply to a fully kinetic calculation. We have attempted to model some of the relevant microphysics by applying hard-wall limits on the pressure anisotropy. This subgrid model is intended to mimic how the mirror and firehose instabilities affect the pressure anisotropy.   But it is unclear whether the limiters are a sufficient and/or accurate representation of the key microphysics. One can also speculate about the relevance of isotropic diffusivities in a collisionless system. Our results suggest that even collisionless-plasma-turbulence may be sensitive to isotropic dissipation. This, then, raises the question of what sets the isotropic diffusivities in a collisionless plasma, where transport properties generally depend on the Larmor radii of the plasma particles and kinetic  microinstabilities that can readily modify the collisionality of the system. To understand this, fully kinetic particle-in-cell simulations are needed. However, such simulations are computationally expensive, and it is unclear whether the simulations can reach a sufficient dynamic range between the outer scale (e.g., disc rotation frequency) and microscopic plasma scales (e.g., the ion cyclotron frequency) to capture the fully developed turbulence that suppresses the pressure anisotropy and anisotropic transport in our simulations. 
\label{sec:dis}

\subsection{Implications for RIAFs}

To conclude we briefly discuss some of the possible implications of our results for RIAFs, bearing in mind the caveat that it is not yet fully clear how our weakly collisional results apply to collisionless plasmas.   Our results support previous work concluding that low-collisionality accretion discs may be reasonably modeled using fluid approximations (e.g., \citealt{fcgqt17}).   Indeed, in some ways our results strengthen this conclusion by showing that the transport in a weakly collisional plasma is more similar to MHD when MRI-driven turbulence is well resolved.   The primary difference relative to MHD in our simulations is that anisotropic viscous heating is a major ($\gtrsim 50 \%$ at high $\re$; see Figure \ref{fig:heating_vs_Re}) source of heating and must be accounted for when modeling the plasma thermodynamics.   Electron heating by anisotropic viscosity (e.g., \citealt{sqhs07}) is thus a key ingredient for models that separately evolve the electron thermodynamics in order to model the radiation from RIAFs (e.g., \citealt{rtqg17}; \citealt{crnjs18}). However, previous estimates of electron vs. proton heating by anisotropic viscosity (e.g., \citealt{sqhs07}) assumed that the electron (proton) pressure anisotropy was near the electron whistler (mirror) threshold.   This was motivated by the fact that the mean shear drives $\Dp> 0$ (Figure \ref{fig:bbdu_transition}b).  However, we have shown that the turbulent fluctuations largely counteract the mean shear (Figure \ref{fig:bbdu_transition}b) and thus viscous heating has significant contributions from both $\Dp> 0$ and $\Dp< 0$.   This should be taken into account in future estimates of anisotropic viscous heating in RIAFs.

\section*{Acknowledgements}
We thank C. Gammie, F. Foucart, A. Schekochihin and J. Stone for useful discussions, as well as all the members of the horizon collaboration, http://horizon.astro.illinois.edu, for their advice and encouragement.   This work was supported in part by NSF grants AST 13-33612, AST 1715054, and a Simons Investigator award from the Simons Foundation. J.S. would like to acknowledge the support of a Rutherford Discovery Fellowship and the Marsden Fund, administered by the New Zealand Royal Society Te Ap\={a}rangi. M.W.K. was supported by NASA grant NNX17AK63G. This work was also  made possible by computing time granted by UCB on the Savio cluster.







\bibliographystyle{mnras}
\bibliography{braginskii}



\appendix
\section{Enhanced Transport in Braginskii MHD}
\label{sec:mod_transport}
\subsection{Dependence on box aspect ratio }
\label{sec:box}
The most notable departure from MHD behavior came about when testing different box aspect ratios for domains with a net vertical field. In addition to our standard  $4 \times 4 \times 1$ boxes, we also looked at narrower domains with aspect ratios $1 \times 4 \times 1$ and $1 \times \pi \times 1$. Consistent with \cite{bmcrf08}, \cite{pg09} and \cite{ll10}, we have also found that MHD simulations in boxes with $L_x :L_z =1$ and net vertical field show stronger fluctuations with recurring bursts. This has been attributed to such boxes not capturing the fastest growing parasitic modes and overemphasizing the role of channel modes.   

While we discovered no surprising behavior in MHD, anisotropic viscosity produced some surprising results. $L_y = 4$ still showed fairly comparable turbulence in MHD and Braginskii MHD, but this was no longer the case for $L_y = \pi$. As the box becomes narrow enough, e.g. $1 \times \pi \times 1$, the presence of anisotropic viscosity strongly enhances transport, with $\alpha$ increasing monotonically with $\nub$. Both the turbulent energy densities and transport can increase by orders of magnitude for large $\nub$. A natural interpretation is that a high anisotropic viscosity causes some parasitic modes to migrate to longer wavelengths. These are not captured by our narrow boxes, thus allowing MRI modes to drive turbulence at a larger amplitude. For boxes with $4 \times 4 \times 1$, this is no longer the case. Another plausible explanation is that anisotropic viscosity decreases the growth rate of short-wavelength modes and so wider boxes are needed to accommodate long-wavelength modes.

\subsection{Braginskii MHD with isotropic hyperdiffusion}
\label{sec:mod_hyp4}
Our composite MHD--Braginskii MHD simulations with small hyperdiffusion ($\Ref = 1.5 \times 10^8$ and $\Ref = 7.5 \times 10^8$) show significant increases in the Maxwell and Reynolds stresses almost immediately after Braginskii viscosity is introduced ($\ratio \sim 0.1$ stays at a roughly constant level nevertheless, as in Figure \ref{fig:alpha_anis_vs_nu4}). For example, in the simulation with $\Ref = 1.5 \times 10^8$, in which the Braginskii MHD part could be evolved for longer ($\orbit = 100 - 130$), $\Tm$  increases by a factor of a few. There are signs of burst-like behavior, suggesting that it may be related to impeded disruption of channel modes in simulations with hyperdiffusion and Braginskii viscosity. Unfortunately, the large anisotropic viscosity and high resolutions at which this behavior occurs prohibit us from thoroughly testing the origin of the increased transport. Thus, it is unclear to what extent this is a physical effect and not simply a numerical artifact associated with our use of hyperdiffusion operators. This result should therefore be treated with caution, especially since we did not see any clear indications of strongly modified $\Tm$ or $\Tr$ in our high-$\re$, second-order-diffusion simulations.

\section{Test case for restarting MHD flow field with Braginskii viscosity} \label{sec:testmixed}

\begin{figure}
\centering
 \includegraphics[scale = 0.75]{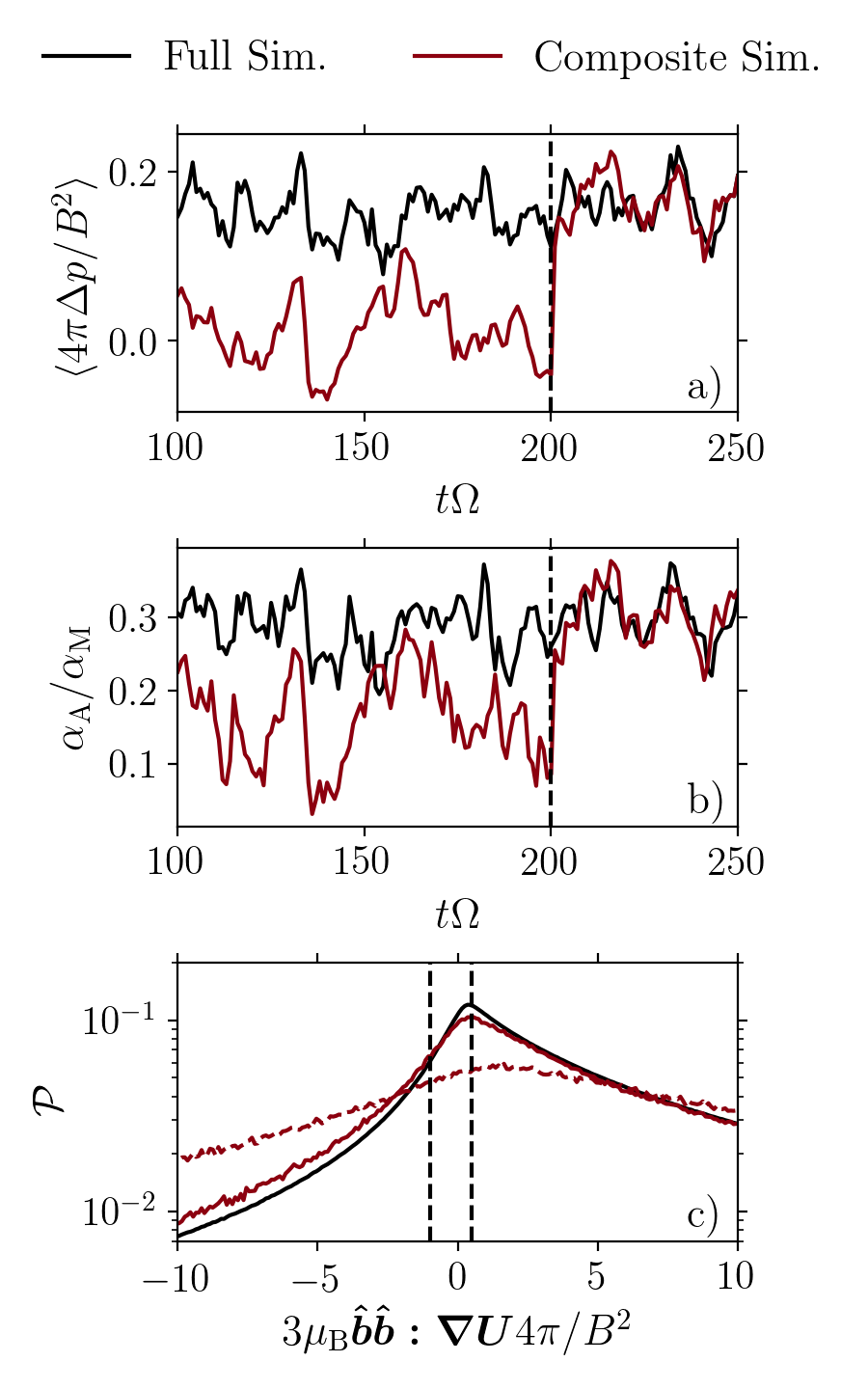}
   \caption{Test of composite MHD--Braginskii MHD simulations for the case $\re=6000$, $\Pm=0.5$ and $\Reb=0.75$. The black line represents the full Braginskii simulation, the crimson line is the composite simulation in which the MHD flow field is restarted with $\Reb=0.75$ at $\orbit = 200$.  \textbf{Panel a):} evolution of $\dpBm$. \textbf{Panel b):} evolution of $\ratio$ (prior to $\orbit=200$, $\Dp$ and $\Ta$ are calculated using the MHD flow field with $\Reb = 0.75$, even though $\nub$ is not present in the MHD equations). \textbf{Panel c):} distributions of $\prel$. The black line is the distribution of the full Braginskii simulation averaged over $\orbit = 100-200$. The dotted crimson line is the MHD distribution at time $\orbit = 200$ and the solid crimson line is the distribution in the composite simulation at $\orbit = 201$, after anisotropic viscosity was introduced. After the MHD flow field is restarted with anisotropic viscosity, the statistics of $\prel$ and the evolution of $\dpBm$ and $\ratio$ agree very well with the full Braginskii simulation.    \label{fig:test_transition} }
\end{figure}

In section \ref{sec:alphaa} we use composite MHD--Braginskii MHD simulations to explore the behavior of anisotropic pressure and stress in the large Reynolds number regime at high resolution. We claimed that by restarting MHD flow fields with anisotropic viscosity, we are able to recover the appropriate Braginskii $\dpm$ in a fraction of an orbital period. This allowed us to determine the anisotropic stress at high Reynolds numbers, which would otherwise be numerically prohibitive with our current methods. Here we justify our ``restart'' method by looking at one of our fully evolved Braginskii MHD simulations.

We focus on the case $\re=6000$, $\Pm=0.5$ and $\Reb = 0.75$. It was evolved for a time $\orbit = 200$ both in MHD and in Braginskii MHD. In Figure \ref{fig:test_transition} we show the evolution of $\dpBm$ and $\ratio$ for the two models. Once again, the pressure anisotropy computed from the MHD flow fields uses the same $\nub$ as the corresponding Braginskii simulation. The plots show that the calculated $\Dp$ is quite different in Braginskii and ordinary MHD, the difference being more significant than the uncertainty related to temporal fluctuations. 

The MHD-generated flow field is then restarted with anisotropic viscosity at time $\orbit = 200$. This induces an almost immediate jump in $\Dp$ and $\Ta$, which can be seen to reach values comparable to the full Braginskii simulation. Afterwards, the two simulations show qualitatively identical evolutions of anisotropic stress and pressure.

Figure \ref{fig:test_transition}c shows how the restarting affects the pre-limiter $\Dp$ distribution. The dashed crimson line is the MHD distribution, immediately before the anisotropic viscosity is introduced at $\orbit  = 200$. The black line is the average distribution of the full Braginskii simulation. The solid crimson curve is the distribution generated by anisotropic viscosity by the time $\orbit = 201$, so just one $\Omega^{-1}$ after the MHD field is restarted. The solid crimson and black curves are in very good agreement with one another, supporting our claim that anisotropic viscosity does not need much time to generate its desired $\bbdU$ distribution. This is not surprising given that the viscous time for $\nub$ in the vertical direction is $\sim \Reb \Omega^{-1} \sim \Omega^{-1}$.


\bsp	
\label{lastpage}

\end{document}